\documentclass[12pt]{article}
\usepackage{a4wide}
\usepackage{latexsym}
\usepackage{amsmath}
\usepackage{amsfonts}
\usepackage{amscd}
\usepackage{cite}
\usepackage{placeins}

\usepackage{pslatex}
\usepackage{graphicx}
\usepackage[latin1,utf8]{inputenc}
\usepackage[T1]{fontenc}

\usepackage{tcolorbox}
\usepackage{colortbl}

\allowdisplaybreaks

\newcommand{\bq}{\begin{eqnarray}}
\newcommand{\eq}{\end{eqnarray}}
\newcommand{\eps}{\varepsilon}

\newcommand{\curveone}{(a)}
\newcommand{\curvetwo}{(b)}

\begin{document}

\thispagestyle{empty}

\begin{flushright}
  MITP/22-037
\end{flushright}

\vspace{1.5cm}

\begin{center}
  {\Large\bf A Feynman integral depending on two elliptic curves\\
  }
  \vspace{1cm}
  {\large Hildegard M\"uller and Stefan Weinzierl \\
  \vspace{1cm}
      {\small \em PRISMA Cluster of Excellence, Institut f{\"u}r Physik, }\\
      {\small \em Johannes Gutenberg-Universit{\"a}t Mainz,}\\
      {\small \em D - 55099 Mainz, Germany}\\
  } 
\end{center}

\vspace{2cm}

\begin{abstract}\noindent
  {
We study a two-loop four-point function with one internal mass.
This Feynman integral is one of the simplest Feynman integrals depending on two elliptic curves.
We transform the associated differential equation into an $\varepsilon$-form.
We study the entries of the differential equation, and in particular the entries which depend
on both elliptic curves.
   }
\end{abstract}

\vspace*{\fill}

\newpage

\section{Introduction}
\label{sect:intro}

Feynman integrals evaluate generically to transcendental functions, in the simplest case to multiple polylogarithms.
Starting from two-loops, we encounter Feynman integrals, which are related to elliptic 
curves \cite{Sabry:1962,Broadhurst:1993mw,Laporta:2004rb,Bailey:2008ib,MullerStach:2011ru,Adams:2013nia,Bloch:2013tra,Remiddi:2013joa,Adams:2014vja,Adams:2015gva,Adams:2015ydq,Bloch:2016izu,Adams:2017ejb,Bogner:2017vim,Adams:2018yfj,Honemann:2018mrb,Bloch:2014qca,Sogaard:2014jla,Tancredi:2015pta,Primo:2016ebd,Remiddi:2016gno,Adams:2016xah,Bonciani:2016qxi,vonManteuffel:2017hms,Adams:2017tga,Ablinger:2017bjx,Primo:2017ipr,Passarino:2017EPJC,Remiddi:2017har,Bourjaily:2017bsb,Hidding:2017jkk,Broedel:2017kkb,Broedel:2017siw,Broedel:2018iwv,Lee:2017qql,Lee:2018ojn,Adams:2018bsn,Adams:2018kez,Broedel:2018qkq,Bourjaily:2018yfy,Bourjaily:2018aeq,Mastrolia:2018uzb,Ablinger:2018zwz,Frellesvig:2019kgj,Broedel:2019hyg,Blumlein:2019svg,Broedel:2019tlz,Bogner:2019lfa,Kniehl:2019vwr,Broedel:2019kmn,Abreu:2019fgk,Duhr:2019rrs,2019arXiv190811815L,Klemm:2019dbm,Bonisch:2020qmm,Weinzierl:2020fyx,Walden:2020odh,Bezuglov:2020ywm,Kristensson:2021ani,Frellesvig:2021hkr}
and go beyond the class of multiple polylogarithms.
Known examples of these types of Feynman integrals evaluate in the univariate case to iterated integrals of modular forms
and in the multivariate case to iterated integrals of integrands related to the coefficients of the Kronecker
function (and modular forms).
It is known that in even more complicated cases the elliptic curve generalises to Calabi-Yau 
manifolds
\cite{Aluffi:2008sy,Brown:2010a,Bourjaily:2019hmc,Vergu:2020uur,Bonisch:2021yfw,Broedel:2021zij},
with an elliptic curve being a Calabi-Yau one-fold.

It is standard practice to compute the Feynman integrals within dimensional regularisation.
A Feynman integral evaluates then to a Laurent series in the dimensional regularisation parameter $\eps$.
We are interested in the coefficients of this Laurent series.
These are the transcendental functions mentioned above.
A convenient tool is the method of differential equations \cite{Kotikov:1990kg,Kotikov:1991pm}.
In particular, if the differential equation for a system of Feynman integrals can be transformed to an
$\eps$-form \cite{Henn:2013pwa}, the Laurent series in the dimensional regularisation parameter $\eps$
follows immediately.
A transformation of the differential equation to an $\eps$-form has been achieved for many Feynman integrals
evaluating to multiple polylogarithms and has been been the driving force 
for the tremendous progress in this field in recent years.
Furthermore, there are examples of Feynman integrals, which depend on a single elliptic curve and where the
associated differential equation has been transformed to an $\eps$-form. 
These integrals can then be solved systematically to all orders in the dimensional regularisation parameter.
For Feynman integrals associated to generic Calabi-Yau manifolds usually only the first non-trivial term in
the $\eps$-expansion has been investigated by other methods and techniques to transform the differential
equation to an $\eps$-form are still missing.

In this paper we go beyond the case of Feynman integrals depending on a single elliptic curve and consider
a two-loop four-point function associated to two elliptic curves.
Throughout this paper we label the two curves by curve $\curveone$ and curve $\curvetwo$.
The corresponding family of Feynman integrals has $12$ master integrals and depends on $2$ kinematic 
variables.
We transform the differential equation to an $\eps$-form and study the entries of the connection matrix.
The entries are differential one-forms.
We will encounter some old friends, already known from multiple polylogarithms and the case of a single elliptic curve: 
dlog-forms, modular forms and one-forms related to the coefficients of
the Kronecker function.
In addition, there are new differential one-forms, depending on both elliptic curves.
These are the objects that will be of particular interest in this paper.

Let us motivate the particular choice of the Feynman integral of this paper: Our aim is to learn something about
Feynman integrals depending on more than one elliptic curve.
To this aim we study one of the simplest examples in this class: 
a two-loop four-point function with one non-zero internal mass, dubbed ``sector $79$''.
This Feynman integral is a sub-topology of the double-box integral with an internal top-loop relevant to
top-pair production at the LHC.
While the latter family of Feynman integrals has $44$ master integrals and depends on three elliptic curves,
the Feynman integral studied in this paper is simpler, while having all essential features: 
The family of Feynman integrals has $12$ master integrals
and depends on two elliptic curves.

We expect that the entries of the differential equation will also show up in other systems of Feynman integrals.
In this sense the differential one-forms appearing in the differential equation are more universal than the specific
Feynman integral we are considering.
For this reason our emphasis is on the entries of the differential equation.

This paper is organised as follows:
In the next section we define the Feynman integral, introduce various kinematic variables and review the Kronecker function.
In section~\ref{sect:master_integrals} we define the master integrals.
In section~\ref{sect:non-mixed} we discuss the entries of the differential equation, which at most depend only on one
elliptic curve. We call these the non-mixed entries.
There are four entries of the differential equation, which depend on both elliptic curves.
These are discussed in section~\ref{sect:mixed}.
Our conclusions are given in section~\ref{sect:conclusions}.
There are two appendices:
Appendix~\ref{sect:mixed_xy} gives the expressions for the mixed entries in the $(x,y)$-coordinates.
(These coordinates are defined in section~\ref{sect:notation}).
In appendix~\ref{sect:dlog_forms} we relate dlog-forms to differential one-forms related to curve $\curvetwo$.


\section{Notation and definitions}
\label{sect:notation}

\subsection{Definition of the Feynman integral}

\begin{figure}
\begin{center}
\includegraphics[scale=1.0]{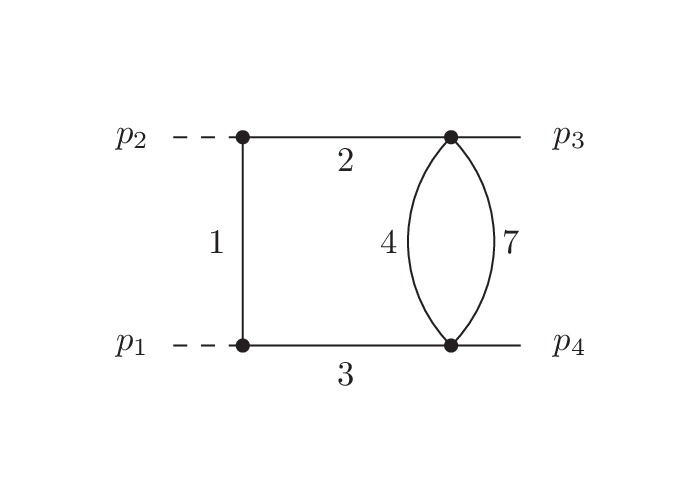}
\end{center}
\caption{
The graph for sector 79.
Solid lines correspond to particles of mass $m$, dashed lines to massless particles.
}
\label{fig_sector_79}
\end{figure}

\begin{figure}
\begin{center}
\includegraphics[scale=1.0]{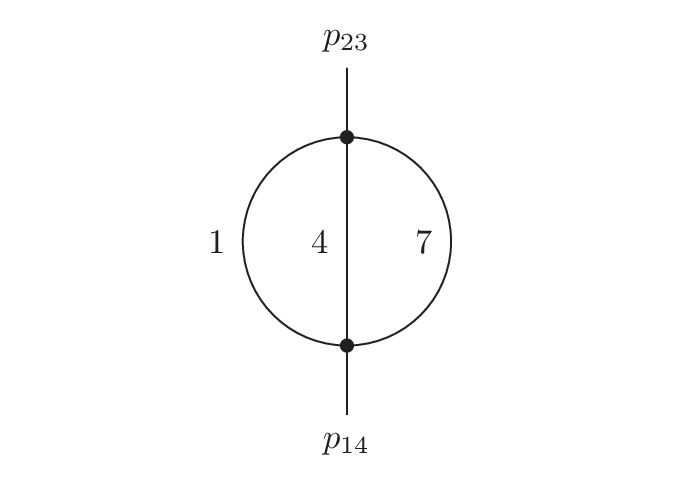}
\end{center}
\caption{
The graph for sector 73.
This graph is known as the sunrise graph. It is obtained by contracting lines $2$ and $3$ from the graph for sector 79. 
}
\label{fig_sector_73}
\end{figure}

We consider the family of two-loop integrals corresponding to the graph shown in fig.~\ref{fig_sector_79}.
The solid internal lines correspond to propagators with a mass $m$.
The external momenta satisfy
\bq
 p_1 + p_2 + p_3 + p_4 = 0,
 & & 
 p_1^2=p_2^2=0, \;\;\;\;\;\; p_3^2=p_4^2=m^2.
\eq
We set
\bq
 s = \left(p_1+p_2\right)^2,
 & &
 t = \left(p_2+p_3\right)^2.
\eq
The graph of fig.~\ref{fig_sector_79} is of particular interest as it is one of the simplest examples where the corresponding
family of Feynman integrals is associated with two elliptic curves:
There is one elliptic curve associated to the graph of fig.~\ref{fig_sector_79} and a second elliptic curve to the sub-graph
shown in fig.~(\ref{fig_sector_73}).
The latter is known as the sunrise graph.

The graph of fig.~\ref{fig_sector_79} occurs as a sub-graph of the planar double box graph relevant to top-pair production
with a closed top loop \cite{Adams:2018bsn,Adams:2018kez}.
It is convenient to adapt the notation from the planar double box graph.
We define a family of Feynman integrals in $D$-dimensional Minkowski space by
\bq
\label{def_integral}
 I_{\nu_1 \nu_2 \nu_3 \nu_4 \nu_5 \nu_6 \nu_7 \nu_8 \nu_9}\left( D, s, t, m^2, \mu^2 \right)
 & = &
 e^{2 \gamma_E \eps}
 \left(\mu^2\right)^{\nu-D}
 \int \frac{d^Dk_1}{i \pi^{\frac{D}{2}}} \frac{d^Dk_2}{i \pi^{\frac{D}{2}}}
 \prod\limits_{j=1}^9 \frac{1}{ P_j^{\nu_j} },
\eq
where $\gamma_E$ denotes the Euler-Mascheroni constant, 
$\mu$ is an arbitrary scale introduced to render the Feynman integral dimensionless, 
the quantity $\nu$ is given by
\bq
 \nu & = &
 \sum\limits_{j=1}^9 \nu_j
\eq
and
\begin{align}
 P_1 & = -\left(k_1+p_2\right)^2 + m^2,
 &
 P_2 & = -k_1^2 + m^2,
 &
 P_3 & = -\left(k_1+p_1+p_2\right)^2 + m^2,
 \nonumber \\
 P_4 & = -\left(k_1+k_2\right)^2 + m^2,
 &
 P_5 & = -k_2^2,
 &
 P_6 & = -\left(k_2+p_3+p_4\right)^2,
 \nonumber \\
 P_7 & = -\left(k_2+p_3\right)^2 + m^2,
 &
 P_8 & = -\left(k_1+p_2-p_3\right)^2 + m^2,
 &
 P_9 & = -\left(k_2-p_2+p_3\right)^2.
\end{align}
A sector is defined by the set of propagators with positive exponents.
We define a sector ID by
\bq
\label{def_sector_id}
 \mathrm{ID}
 & = & \sum\limits_{j=1}^9 2^{j-1} \Theta(\nu_j).
\eq
In this article we are interested in the sector $79$ and all of its subsectors.
This is the subset of Feynman integrals, where none of the propagators $\{P_5,P_6,P_8,P_9\}$ has a positive
exponent.
The relevant topologies are shown in section~\ref{sect:master_integrals} in fig.~\ref{fig_master_topologies}.

We define the dimensional shift operators ${\bf D}^\pm$ by
\bq
\label{def_dimensional_shift}
 {\bf D}^\pm I_{\nu_1 \nu_2 \nu_3 \nu_4 \nu_5 \nu_6 \nu_7 \nu_8 \nu_9}\left( D \right)
 & = &
 I_{\nu_1 \nu_2 \nu_3 \nu_4 \nu_5 \nu_6 \nu_7 \nu_8 \nu_9}\left( D \pm 2 \right).
\eq

\subsection{Coordinates}

Without loss of generality we may set $\mu=m$ in
eq.~(\ref{def_integral}).
Then the Feynman integrals in eq.~(\ref{def_integral})
depend only on two dimensionless ratios, which may be taken as
\bq
\label{s_t_coordinates}
 \frac{s}{m^2},
 & &
 \frac{t}{m^2}.
\eq
In other words, we may view the integrals $I_{\nu_1 \nu_2 \nu_3 \nu_4 \nu_5 \nu_6 \nu_7 \nu_8 \nu_9}$ as functions
on $M={\mathbb P}^2({\mathbb C})$, where
\bq
 \left[ s : t : m^2 \right]
\eq
denote the homogeneous coordinates.

Note that we are free to choose any convenient coordinates on $M$.
One possibility is given by eq.~(\ref{s_t_coordinates}).
We will refer to this choice as $(\frac{s}{m^2},\frac{t}{m^2})$-coordinates.
Other choices which we use are
\bq
 \left(x,y\right),
 \;\;\;
 \left(x',y'\right),
 \;\;\;
 \left(\tau^{\curveone},\tau^{\curvetwo}\right),
 \;\;\;
 \left(\bar{q}^{\curveone},\bar{q}^{\curvetwo}\right),
 \;\;\;
 \left(z^{\curvetwo},\tau^{\curvetwo}\right),
 \;\;\;
 \left(\bar{w}^{\curvetwo},\bar{q}^{\curvetwo}\right).
\eq
We will move frequently between different choices of coordinates.
With the exception of $z^{\curvetwo}$ and $\bar{w}^{\curvetwo}$ we define the coordinates below.
The coordinates $z^{\curvetwo}$ and $\bar{w}^{\curvetwo}$ are defined in section~\ref{sect:constructing_z}.
The reason for postponing the definition of $z^{\curvetwo}$ and $\bar{w}^{\curvetwo}$ is the following: 
The defining equation for $z^{\curvetwo}$ (and in turn $\bar{w}^{\curvetwo}$) is obtained from the differential equation 
of the master integrals. 
Thus we first define the master integrals in section~\ref{sect:master_integrals} and the coordinates $z^{\curvetwo}$ and $\bar{w}^{\curvetwo}$
afterwards.

Let us now consider the coordinates $x$, $y$, $x'$, $y'$, $\tau^{\curveone}$, $\tau^{\curvetwo}$, $\bar{q}^{\curveone}$ and
 $\bar{q}^{\curvetwo}$.
We start with the coordinate set $(x,y)$.
The coordinates $x$ and $y$ are related to $s$ and $t$ by
\bq
 \frac{s}{m^2} \; = \; - \frac{\left(1-x\right)^2}{x},
 & &
 \frac{t}{m^2} \; = \; y.
\eq
We will refer to this choice as $(x,y)$-coordinates.
The coordinate $x$ rationalises the square root $\sqrt{-s(4m^2-s)}$.
For the inverse transformation we choose the sign such that $s=-\infty$ corresponds to $x=0$:
\bq
 x & = &
 \frac{1}{2} \left[ \frac{-s}{m^2} + 2 - \frac{1}{m^2} \sqrt{ -s \left( 4 m^2 - s \right)}\right].
\eq
In addition it will be convenient to rationalise the zeros of the two quartic polynomials defining the two elliptic curves
(defined below in eq.~(\ref{zeros_curveone}) and eq.~(\ref{zeros_curvetwo})).
We encounter the two roots
\bq
 \sqrt{\frac{t}{m^2}},
 & &
 \sqrt{\frac{t}{m^2} + \frac{\left(m^2-t\right)^2}{m^2 s}}.
\eq
This can be done by \cite{Besier:2018jen,Besier:2019kco}
\bq
 x \; = \; - \frac{\left[\left(1-x'\right)^2 -y'^4 \left(1+x'\right)^2\right]}{4 x' y'^2},
 & &
 y \; = \; y'^2.
\eq
We will refer to this choice as $(x',y')$-coordinates.
The inverse transformation is given by
\bq
 x' \; = \;
 \frac{1+y^2-2xy-2\sqrt{y\left(y-x\right)\left(1-xy\right)}}{1-y^2},
 & &
 y' \; = \; \sqrt{y}
\eq
and maps $x=0$ to $x'=(1-y)/(1+y)$.

Derivatives with respect to one coordinate are taken within a coordinate chart with the other coordinates kept
fixed.
Thus
\bq
 \frac{\partial f}{\partial y}
\eq
denotes the derivative of $f$ with respect to $y$ with $x$ kept fixed, while
\bq
 \frac{\partial f}{\partial y'}
\eq
denotes the derivative of $f$ with respect to $y'$ with $x'$ kept fixed.
As $\partial y' / \partial x = 0$ we have
\bq
\label{chain_rule}
 \frac{\partial f}{\partial x}
 & = &
 \left( \frac{\partial x'}{\partial x} \right) \frac{\partial f}{\partial x'},
 \nonumber \\
 \frac{\partial f}{\partial y}
 & = &
 \left( \frac{\partial x'}{\partial y} \right) \frac{\partial f}{\partial x'}
 +
 \left( \frac{\partial y'}{\partial y} \right) \frac{\partial f}{\partial y'}.
\eq
There are two more sets of coordinates, which we are going to use: These are the sets $(\tau^{\curveone},\tau^{\curvetwo})$ and
$(\bar{q}^{\curveone},\bar{q}^{\curvetwo})$ where the superscripts $\curveone$ and $\curvetwo$ refer to the two elliptic curves. These coordinates are obtained as follows:

We start from a general elliptic curve defined by a quartic polynomial
\bq
\label{def_generic_quartic_elliptic_curve}
 E
 & : &
 w^2 - \left(z-z_1\right) \left(z-z_2\right) \left(z-z_3\right) \left(z-z_4\right)
 \; = \; 0,
\eq
where the $z_j$ (with $j\in\{1,2,3,4\}$) denote the roots of the quartic polynomial.
We set
\bq
\label{def_Z}
 Z_1 \; = \; \left(z_3-z_2\right)\left(z_4-z_1\right),
 \;\;\;\;\;\;
 Z_2 \; = \; \left(z_2-z_1\right)\left(z_4-z_3\right),
 \;\;\;\;\;\;
 Z_3 \; = \; \left(z_3-z_1\right)\left(z_4-z_2\right).
\eq 
We define the modulus and the complementary modulus of the elliptic curve $E$ by
\bq
 k^2 
 \; = \; 
 \frac{Z_1}{Z_3},
 & &
 \bar{k}^2 
 \; = \;
 1 - k^2 
 \; = \;
 \frac{Z_2}{Z_3}.
\eq
Our standard choice for the periods and quasi-periods is
\bq
\label{def_generic_periods}
 \psi_1 
 \; = \; 
 \frac{4 K\left(k\right)}{Z_3^{\frac{1}{2}}},
 & &
 \psi_2
 \; = \; 
 \frac{4 i K\left(\bar{k}\right)}{Z_3^{\frac{1}{2}}},
 \nonumber \\
 \phi_1 
 \; = \; 
 \frac{4 \left[ K\left(k\right) - E\left(k\right) \right]}{Z_3^{\frac{1}{2}}},
 & &
 \phi_2
 \; = \; 
 \frac{4 i E\left(\bar{k}\right)}{Z_3^{\frac{1}{2}}}.
\eq
The two elliptic curves $E^{\curveone}$ and $E^{\curvetwo}$ are obtained from the maximal cut 
of the corresponding Feynman integrals \cite{Adams:2018bsn,Adams:2018kez}. To define the elliptic curves we just have to give the four roots $z_1$-$z_4$.
For the elliptic curve $E^{\curveone}$ we specialise to
\bq
\label{zeros_curveone}
 z^{\curveone}_1
 \; = \;
 y'^2-4,
 \;\;\;\;\;\;
 z^{\curveone}_2
 \; = \;
 -1-2y',
 \;\;\;\;\;\;
 z^{\curveone}_3
 \; = \;
 -1+2y',
 \;\;\;\;\;\;
 z^{\curveone}_4
 \; = \;
 y'^2,
\eq
for the elliptic curve $E^{\curvetwo}$ we specialise to 
\bq
\label{zeros_curvetwo}
 & &
 z^{\curvetwo}_1
 \; = \;
 y'^2-4,
 \;\;\;\;\;\;
 z^{\curvetwo}_2
 \; = \;
 -1-2\chi^{\curvetwo},
 \;\;\;\;\;\;
 z^{\curvetwo}_3
 \; = \;
 -1+2\chi^{\curvetwo},
 \;\;\;\;\;\;
 z^{\curvetwo}_4
 \; = \;
 y'^2,
\eq
with
\bq
\label{def_chi}
 \chi^{\curvetwo}
 & = &
 \frac{y'\left(1+y'^2\right)\left(1-x'^2\right)}{\left[\left(1-x'\right)^2+y'^2\left(1+x'\right)^2\right]}.
\eq
Eq.~(\ref{zeros_curveone}) together with eq.~(\ref{def_generic_quartic_elliptic_curve}) defines the elliptic curve $\curveone$,
eq.~(\ref{zeros_curvetwo}) together with eq.~(\ref{def_generic_quartic_elliptic_curve}) defines the elliptic curve $\curvetwo$.
Furthermore, eq.~(\ref{def_generic_periods}) defines then the periods and quasi-periods for the curve $(a)$ and $(b)$:
\bq
 \psi_1^{\curveone},
 \;
 \psi_2^{\curveone}, 
 \;
 \phi_1^{\curveone},
 \;
 \phi_2^{\curveone}, 
 \;
 \psi_1^{\curvetwo},
 \;
 \psi_2^{\curvetwo}, 
 \;
 \phi_1^{\curvetwo},
 \;
 \phi_2^{\curvetwo}.
\eq
The Wronskians are defined by
\bq
 W^{(c)}_z 
 & = &
 \psi_1^{(c)} \frac{d}{dz} \psi_2^{(c)}
 -
 \psi_2^{(c)} \frac{d}{dz} \psi_1^{(c)}
 \; = \;
 \frac{4\pi i}{Z_3^{(c)}} \frac{d}{dz} \ln \frac{Z_2^{(c)}}{Z_1^{(c)}}
\eq
with
\bq
 c \; \in \; \left\{a,b\right\},
 & & 
 z \; \in \; \left\{ x,y,x',y' \right\}.
\eq
We have for example
\bq
 W^{\curveone}_y
 & = &
 -\frac{6 \pi i}{y\left(1-y\right)\left(9-y\right)},
 \nonumber \\
 W^{\curvetwo}_y
 & = &
 -\frac{2 \pi i \left(1-x\right)^2 \left(3x^2-2xy-4x+3\right)}{\left(1-y\right)\left(y-x\right)\left(1-xy\right)\left(9-14x-y-2xy+9x^2-x^2y\right)}.
\eq
We further define the modular parameters
\bq
\label{def_tau}
 \tau^{\curveone} \; = \; \frac{\psi_2^{\curveone}}{\psi_1^{\curveone}},
  & &
 \tau^{\curvetwo} \; = \; \frac{\psi_2^{\curvetwo}}{\psi_1^{\curvetwo}},
\eq
and
\bq
 \bar{q}^{\curveone} \; = \; e^{2 \pi i \tau^{\curveone}},
 & &
 \bar{q}^{\curvetwo} \; = \; e^{2 \pi i \tau^{\curvetwo}}.
\eq
In addition we set
\bq
 \tau^{\curveone}_n \; = \; \frac{\tau^{\curveone}}{n},
 & &
 \bar{q}^{\curveone}_n \; = \; e^{2 \pi i \tau^{\curveone}_n} \; = \; e^{\frac{2 \pi i \tau^{\curveone}}{n}},
 \nonumber \\
 \tau^{\curvetwo}_n \; = \; \frac{\tau^{\curvetwo}}{n},
 & &
 \bar{q}^{\curvetwo}_n \; = \; e^{2 \pi i \tau^{\curvetwo}_n} \; = \; e^{\frac{2 \pi i \tau^{\curvetwo}}{n}}.
\eq
The equations above define $\bar{q}^{\curveone}_2$ and $\bar{q}^{\curvetwo}_2$
as functions of $x'$ and $y'$.
We would like to change variables from $(x',y')$ to $(\bar{q}^{\curveone}_2,\bar{q}^{\curvetwo}_2)$.
This requires the inverse relation:
We would like to express the variables $x'$ and $y'$ in terms of the 
variables $\bar{q}^{\curveone}_2$ and $\bar{q}^{\curvetwo}_2$. This can be done as follows:
We may exchange the variable $y'$ with $\bar{q}^{\curveone}_2$:
\bq
\label{hauptmodul_0}
 y'
 & = &
 3
 \frac{\eta\left(2\tau^{\curveone}_2\right)^2 \eta\left(12\tau^{\curveone}_2\right)^4}
      {\eta\left(6\tau^{\curveone}_2\right)^2 \eta\left(4\tau^{\curveone}_2\right)^4},
\eq
where $\eta$ denotes Dedekind's eta-function.
The first few terms read
\bq
 y' & = & 
 3 \bar{q}_2^{\curveone}
 - 6 \left(\bar{q}_2^{\curveone}\right)^3
 + 9 \left(\bar{q}_2^{\curveone}\right)^5
 + {\mathcal O}\left(\left(\bar{q}_2^{\curveone}\right)^7\right).
\eq
This change of variables is known from the literature \cite{Maier:2006aa,Bloch:2013tra,Adams:2017ejb},
our notation follows closely ref.~\cite{Honemann:2018mrb}.
In order to express the variable $x'$ in $\bar{q}^{\curveone}_2$ and $\bar{q}^{\curvetwo}_2$ we proceed as follows:
We consider 
\bq
 \Lambda & = & \frac{1}{16} \left( k^{\curvetwo} \right)^2 \left( \bar{k}^{\curvetwo} \right)^2.
\eq
On the one hand $\Lambda$ is a rational function in $x'$ and $y'$
\bq
 \Lambda & = & \frac{Z_1^{\curvetwo} Z_2^{\curvetwo}}{16 \left(Z_3^{\curvetwo}\right)^2},
\eq
which has a Taylor expansion around $x'=1$, starting with the linear term.
On the other hand we have
\bq
 \Lambda
 & = &
 \frac{\eta\left(\tau^{\curvetwo}_2\right)^{24}\eta\left(4\tau^{\curvetwo}_2\right)^{24}}{\eta\left(2\tau^{\curvetwo}_2\right)^{48}}.
\eq
Here we may view $\Lambda$ as a power series in $\bar{q}^{\curvetwo}_2$, again starting with the linear term.
Power series reversion gives then $x'$ as a power series in $\bar{q}^{\curvetwo}_2$ with coefficients being
functions of $y'$:
\bq
 x' & = &
 1
 - 2 y' \left(3-y'^2\right) \bar{q}_2^{\curvetwo}
 + 2 y'^2 \left(3-y'^2\right)^2 \left(\bar{q}_2^{\curvetwo}\right)^2
 \nonumber \\
 & &
 - 2 y' \left(3-y'^2\right) \left( 1 + 11 y'^2 - 9 y'^4 + y'^6 \right) \left( \bar{q}_2^{\curvetwo} \right)^3
 \nonumber \\
 & &
 + 2 y'^2 \left(3-y'^2\right)^2 \left( 2 + 13 y'^2 - 12 y'^4 + y'^6 \right) \left(\bar{q}_2^{\curvetwo}\right)^4
 + {\mathcal O}\left(\left(\bar{q}_2^{\curvetwo}\right)^5\right).
\eq
For $y'$ we then use eq.~(\ref{hauptmodul_0}).
This gives us a power series in $\bar{q}^{\curveone}_2$ and $\bar{q}^{\curvetwo}_2$.

\subsection{Singularities}
\label{sect:singularities}

It is helpful to know the singularities of the differential equation.
These are already known from refs.~\cite{Adams:2018bsn,Adams:2018kez}.
In the denominators we encounter the following polynomials (normalised with $m^2$):
Polynomials, which only depend on $x$ are
\bq
 \frac{\left(-s\right)}{m^2} 
 \; = \; 
 \frac{\left(1-x\right)^2}{x},
 & &
 \frac{4m^2-s}{m^2}
 \; = \;
 \frac{\left(1+x\right)^2}{x}.
\eq
Polynomials, which only depend on $y$ are
\bq
 \frac{t}{m^2} 
 \; = \; 
 y,
 \;\;\;\;\;\;\;\;\;
 \frac{m^2-t}{m^2} 
 \; = \; 
 1-y,
 \;\;\;\;\;\;\;\;\;
 \frac{9m^2-t}{m^2} 
 \; = \; 
 9-y.
\eq
Polynomials, which depend on $x$ and $y$ are
\bq
 \frac{st +\left(m^2-t\right)^2}{m^4}
 & = & 
 - \frac{\left(1-xy\right)\left(y-x\right)}{x},
 \nonumber \\
 \frac{m^2-t-s}{m^2}
 & = & 
 \frac{1-x+x^2-xy}{x},
 \nonumber \\
 \frac{3s+2t-2m^2}{m^2}
 & = & 
 - \frac{3-4x+3x^2-2xy}{x},
 \nonumber \\
 \frac{s\left(t-9m^2\right)+4m^2\left(m^2-t\right)}{m^4}
 & = & 
 \frac{9-14x-y-2xy+9x^2-x^2y}{x}.
\eq
At the point $x=y=0$ the polynomials
\bq
 x,
 \;\;\;\;\;\;
 y,
 \;\;\;\;\;\;
 y-x
\eq
vanish, all other polynomials attain a finite non-zero value.

\subsection{Special kinematic configurations}
\label{sect:special_kinematic_configuration}

Let us discuss a few special kinematic configurations:
For $y=0$ we have 
\bq
 y \; = \; 0
 & : &
 y' \; = \; 0,
 \;\;\;\;\;\;
 \tau^{\curveone} \; = \; i \infty,
 \;\;\;\;\;\;
 \bar{q}^{\curveone} \; = \; 0.
\eq
In this limit the two roots $z^{\curveone}_2$ and $z^{\curveone}_3$ coincide and the
curve $\curveone$ degenerates to a nodal curve.

For $x=y$ we have 
\bq
 x \; = \; y
 & : &
 x' \; = \; 1,
 \;\;\;\;\;\;
 \tau^{\curvetwo} \; = \; i \infty,
 \;\;\;\;\;\;
 \bar{q}^{\curvetwo} \; = \; 0.
\eq
In this limit the two roots $z^{\curvetwo}_2$ and $z^{\curvetwo}_3$ coincide and the
curve $\curvetwo$ degenerates to a nodal curve.

The third special case is the case $x=0$, for which we have
\bq
 x \; = \; 0
 & : &
 x' \; = \; \frac{1-y'^2}{1+y'^2},
 \;\;\;\;\;\;
 \tau^{\curveone} \; = \; \tau^{\curvetwo},
 \;\;\;\;\;\;
 \bar{q}^{\curveone} \; = \; \bar{q}^{\curvetwo}.
\eq
We are interested in the behaviour of the differential equation in these limits.
The study of these limits gives us useful information.
The limits are simpler, as they only depend on one kinematic variable.
However, in these limits we have to treat the differential one-forms
\bq
 d\ln\left(x\right),
 \;\;\;\;\;\;
 d\ln\left(y\right),
 \;\;\;\;\;\;
 d\ln\left(y-x\right)
\eq
carefully.
This is of particular importance, as we would like to change freely between different coordinate systems. 
We proceed as follows: For the limit $x\rightarrow 0$ 
we first subtract from a differential one-form of interest a suitable multiple of
$d\ln(x)$, such that the difference is regular at $x=0$.
We then study the regular remainder.
The limits $y\rightarrow 0$ and $x\rightarrow y$ are treated in a similar way.

This procedure avoids the following problems:

\begin{enumerate}

\item Consider the limit $x=0$. The case $x=0$ implies $x=\mathrm{const}$ and we would like to set all differential
one-forms proportional to $dx$ to zero. This includes $dx/x$.
Consider now a variable transformation $\tilde{x}=y x$ and suppose $y \neq 0$.
The points $(x,y)=(0,y)$ in a neighbourhood of $y_0\neq0$ 
are mapped to $(\tilde{x},y)=(0,y)$, hence we have $\tilde{x}=\mathrm{const}$.
Setting $d\tilde{x}/{\tilde{x}}$ to zero leads to the contradiction
\bq 
 d\ln\left(\tilde{x}\right) \; = \; d\ln\left(x\right) + d\ln\left(y\right)
 & \Rightarrow &
 d\ln\left(y\right) \; = \; 0.
\eq
Of course, the problem originates from the fact that $dx/x$ and $d\tilde{x}/\tilde{x}$ are actually of the type $0/0$.

\item We consider again the limit $x=0$, implying $\bar{q}^{\curveone}=\bar{q}^{\curvetwo}$.
Very often we will first expand in $\bar{q}^{\curvetwo}$ followed by an expansion in $\bar{q}^{\curveone}$.
For this double expansion we will assume the hierarchy
\bq
 \left| \bar{q}^{\curvetwo} \right| \; < \; \left| \bar{q}^{\curveone} \right| \; \ll \; 1,
\eq
such that $|\frac{\bar{q}^{\curvetwo}}{\bar{q}^{\curveone}}|<1$.
This includes a neighbourhood of $x=y$, but does not include the region $x=0$.
To give an example, in a neighbourhood of $x=y$ we may expand
the following logarithm in $\bar{q}^{\curvetwo}$:
\bq
 \ln\left(\bar{q}^{\curveone}-\bar{q}^{\curvetwo}\right)
 & = &
 \ln\left(\bar{q}^{\curveone}\right)
 - \sum\limits_{j=1}^\infty \frac{1}{j} \left( \frac{\bar{q}^{\curvetwo}}{\bar{q}^{\curveone}} \right)^j.
\eq
Obviously, the series does not converge for $\bar{q}^{\curveone}=\bar{q}^{\curvetwo}$.

\end{enumerate}


\subsection{The Kronecker function}
\label{sect:kronecker_function}

In this section we review the Kronecker function $F(z,y,\tau)$.
This function and its expansion in $y$ appear frequently in the elliptic setting.
Standard references are \cite{Zagier:1991,Brown:2011,Broedel:2018qkq}.
Apart from the well-known expansion around $y=0$ we also introduce the expansion around $y=\frac{1}{2}$.

The Kronecker function $F(z,y,\tau)$ is defined in terms of the first Jacobi theta function by
\bq
 F\left(z,y,\tau\right)
 & = &
 \bar{\theta}_1'\left(0,\bar{q}\right) \frac{\bar{\theta}_1\left(z+y, \bar{q}\right)}{\bar{\theta}_1\left(z, \bar{q}\right)\bar{\theta}_1\left(y, \bar{q}\right)},
\eq
where 
\bq
 \bar{\theta}_1\left(z,\bar{q}\right) 
 & = &
 -i \sum\limits_{n=-\infty}^\infty \left(-1\right)^n \bar{q}^{\frac{1}{2}\left(n+\frac{1}{2}\right)^2} e^{i \pi\left(2n+1\right)z}.
\eq
$\bar{\theta}_1'$ denotes the derivative with respect to the first argument.
For our purpose a more convenient representation is given by
\bq
 F\left(z,y,\tau\right)
 & = &
 - 2 \pi i 
 \left[
  \frac{1+\bar{w}}{2\left(1-\bar{w}\right)}
  +
  \frac{1+\bar{v}}{2\left(1-\bar{v}\right)}
  +
  \overline{\mathrm{E}}_{0;0}\left(\bar{w};\bar{v};\bar{q}\right)
 \right],
\eq
where $\bar{w}=\exp(2\pi i z)$, $\bar{v}=\exp(2\pi i y)$
and
\bq
 \overline{\mathrm{E}}_{n;m}\left(\bar{w};\bar{v};\bar{q}\right) 
 & = &
  \mathrm{ELi}_{n;m}\left(\bar{w};\bar{v};\bar{q}\right)
  - \left(-1\right)^{n+m} \mathrm{ELi}_{n;m}\left(\bar{w}^{-1};\bar{v}^{-1};\bar{q}\right),
 \nonumber \\
 \mathrm{ELi}_{n;m}\left(\bar{w};\bar{v};\bar{q}\right) & = & 
 \sum\limits_{j=1}^\infty \sum\limits_{k=1}^\infty \; \frac{\bar{w}^j}{j^n} \frac{\bar{v}^k}{k^m} \bar{q}^{j k}.
\eq
The Kronecker function is symmetric in $z$ and $y$.
We are interested in the Laurent expansion in one of these variables. 
We define functions $g^{(k)}(z,\tau)$ through
\bq
\label{def_g_n}
 F\left(z,y,\tau\right)
 & = &
 \sum\limits_{k=0}^\infty g^{(k)}\left(z,\tau\right) y^{k-1}.
\eq
The coefficient functions $g^{(k)}(z,\tau)$ have already appeared in many applications \cite{Zagier:1991,Brown:2011,Broedel:2018qkq}.

In addition to the functions $g^{(k)}(z,\tau)$, which appear in the expansion of the Kronecker function
around $y=0$ we define
functions $h^{(k)}(z,\tau)$, which appear in the expansion of the Kronecker function around $y=\frac{1}{2}$:
\bq
\label{def_h_n}
 F\left(z,y,\tau\right)
 & = &
 \sum\limits_{k=0}^\infty h^{(k)}\left(z,\tau\right) \left(y-\frac{1}{2}\right)^{k-1}.
\eq
The $\bar{q}$-expansion of the $g^{(k)}(z,\tau)$ functions is given by 
\bq
 g^{(0)}\left(z,\tau\right)
 & = & 1,
 \nonumber \\
 g^{(1)}\left(z,\tau\right)
 & = &
 - 2 \pi i \left[
                  \frac{1+\bar{w}}{2 \left(1-\bar{w}\right)}
                  + \overline{\mathrm{E}}_{0,0}\left(\bar{w};1;\bar{q}\right)
 \right],
 \nonumber \\
 g^{(k)}\left(z,\tau\right)
 & = &
 - \frac{\left(2\pi i\right)^k}{\left(k-1\right)!} 
 \left[
 - \frac{B_k}{k}
       + \overline{\mathrm{E}}_{0,1-k}\left(\bar{w};1;\bar{q}\right)
 \right],
 \;\;\;\;\;\;\;\;\;
 k > 1,
\eq
where $B_k$ denotes the $k$-th Bernoulli number, defined by
\bq
 \frac{x}{e^x-1}
 & = &
 \sum\limits_{j=0}^\infty \frac{B_j}{j!} x^j.
\eq
The first few Bernoulli numbers are 
\bq
 B_0 \; = \; 1,
 \;\;\;\;\;\;
 B_1 \; = \; -\frac{1}{2},
 \;\;\;\;\;\;
 B_2 \; = \; \frac{1}{6},
 \;\;\;\;\;\;
 B_3 \; = \; 0,
 \;\;\;\;\;\;
 B_4 \; = \; -\frac{1}{30}.
\eq
The $\bar{q}$-expansion of the $h^{(k)}(z,\tau)$ functions is given by 
\bq
 h^{(0)}\left(z,\tau\right)
 & = & 0,
 \nonumber \\
 h^{(1)}\left(z,\tau\right)
 & = &
 - 2 \pi i \left[
                  \frac{1+\bar{w}}{2 \left(1-\bar{w}\right)}
                  + \overline{\mathrm{E}}_{0,0}\left(\bar{w};-1;\bar{q}\right)
 \right],
 \nonumber \\
 h^{(k)}\left(z,\tau\right)
 & = &
 - \frac{\left(2\pi i\right)^k}{\left(k-1\right)!} 
 \left[
 \left(1-2^k\right) \frac{B_k}{k}
       + \overline{\mathrm{E}}_{0,1-k}\left(\bar{w};-1;\bar{q}\right)
 \right],
 \;\;\;\;\;\;\;\;\;
 k > 1.
\eq
We define the differential one-forms
\bq
\label{def_omega_Kronecker}
 \omega^{\mathrm{Kronecker}}_{k}\left(z,\tau\right)
 & = &
 \left(2\pi i\right)^{2-k}
 \left[
  g^{(k-1)}\left( z, \tau\right) d z + \left(k-1\right) g^{(k)}\left( z, \tau\right) \frac{d\tau}{2\pi i}
 \right],
 \nonumber \\
 \omega^{\mathrm{Kronecker},\frac{1}{2}}_{k}\left(z,\tau\right)
 & = &
 \left(2\pi i\right)^{2-k}
 \left[
  h^{(k-1)}\left( z, \tau\right) d z + \left(k-1\right) h^{(k)}\left( z, \tau\right) \frac{d\tau}{2\pi i}
 \right].
\eq
These differential one-forms are closed
\bq
 d \omega^{\mathrm{Kronecker}}_{k} \; = \; d \omega^{\mathrm{Kronecker},\frac{1}{2}}_{k} & = & 0.
\eq
We will encounter $\omega^{\mathrm{Kronecker},\frac{1}{2}}_{2}$ and $\omega^{\mathrm{Kronecker},\frac{1}{2}}_{3}$.
These may be reduced to $\omega^{\mathrm{Kronecker}}_{2}$ and $\omega^{\mathrm{Kronecker}}_{3}$
with the help of the following formulae
\bq
\label{reduction_omega_Kronecker_onehalf}
 \omega^{\mathrm{Kronecker},\frac{1}{2}}_{2}\left(z,\tau\right)
 & = &
 - \omega^{\mathrm{Kronecker}}_{2}\left(z,\tau\right)
 + 2 \omega^{\mathrm{Kronecker}}_{2}\left(\frac{z}{2},\tau\right)
  + 2 \omega^{\mathrm{Kronecker}}_{2}\left(\frac{z}{2}+\frac{1}{2},\tau\right),
 \nonumber \\
 \omega^{\mathrm{Kronecker},\frac{1}{2}}_{3}\left(z,\tau\right)
 & = &
 - \omega^{\mathrm{Kronecker}}_{3}\left(z,\tau\right)
 + 4 \omega^{\mathrm{Kronecker}}_{3}\left(\frac{z}{2},\tau\right)
  + 4 \omega^{\mathrm{Kronecker}}_{3}\left(\frac{z}{2}+\frac{1}{2},\tau\right).
 \;\;\;
\eq
Eq.~(\ref{reduction_omega_Kronecker_onehalf}) allows us to deduce the modular transformation properties
of $\omega^{\mathrm{Kronecker},\frac{1}{2}}_{2}$ and $\omega^{\mathrm{Kronecker},\frac{1}{2}}_{3}$
from the known modular transformation properties of $\omega^{\mathrm{Kronecker}}_{k}$.
Under a modular transformation the variables $\tau$ and $z$ transform as
\bq
 \tau' \; = \; \frac{a\tau+b}{c\tau+d},
 \;\;\;\;\;\;
 z' \; = \; \frac{z}{c\tau +d},
 \;\;\;\;\;\;
 \left(\begin{array}{cc}
   a & b \\
   c & d \\
 \end{array} \right)
 \; \in \;
 \mathrm{SL}_2\left(\mathbb{Z}\right).
\eq
We have to consider a slight generalisation: Let
\bq
 L\left(z\right) & = & \alpha z + \beta
\eq
be a linear function of $z$ and set
\bq
 \alpha' \; = \; \alpha, 
 & &
 \beta' \; = \; \frac{\beta}{c\tau+d}.
\eq
Then
\bq
 L'\left(z'\right)
 & = & 
 \frac{L\left(z\right)}{c\tau+d}
\eq
and
\bq
\label{trafo_omega_Kronecker}
 \omega^{\mathrm{Kronecker}}_k\left(L'(z'),\tau'\right)
 & = &
 \left(c\tau +d \right)^{k-2}
 \sum\limits_{j=0}^k
 \frac{1}{j!}
 \left( \frac{c L(z)}{c\tau+d} \right)^j
 \omega^{\mathrm{Kronecker}}_{k-j}\left(L(z),\tau\right).
\eq
Eq.~(\ref{reduction_omega_Kronecker_onehalf}) allows us as well to deduce the periodicity and reflection properties, for example
\bq
 \omega^{\mathrm{Kronecker},\frac{1}{2}}_{k}\left(z+1,\tau\right)
 & = &
 \omega^{\mathrm{Kronecker},\frac{1}{2}}_{k}\left(z,\tau\right),
 \nonumber \\
 \omega^{\mathrm{Kronecker},\frac{1}{2}}_{k}\left(-z,\tau\right)
 & = & 
 \left(-1\right)^k \omega^{\mathrm{Kronecker},\frac{1}{2}}_{k}\left(z,\tau\right),
\eq
for $k \in \{2,3\}$.
Eq.~(\ref{reduction_omega_Kronecker_onehalf}) also states that $\omega^{\mathrm{Kronecker},\frac{1}{2}}_{2}$ and $\omega^{\mathrm{Kronecker},\frac{1}{2}}_{3}$ can be eliminated.
However, it is useful to keep $\omega^{\mathrm{Kronecker},\frac{1}{2}}_{2}$ and $\omega^{\mathrm{Kronecker},\frac{1}{2}}_{3}$
as some expressions are shorter when expressed in terms of these one-forms.

Locally we may introduce primitives for $\omega^{\mathrm{Kronecker}}_{k}$.
We set
\bq
 \omega^{\mathrm{Kronecker}}_{k} & = & d \Omega^{\mathrm{Kronecker}}_{k}.
\eq
This defines $\Omega^{\mathrm{Kronecker}}_{k}$ up to an additive constant.
We have
\bq
 \Omega^{\mathrm{Kronecker}}_{0}\left(z,\tau\right)
 & = & 
 -\ln\bar{q},
 \\
 \Omega^{\mathrm{Kronecker}}_{1}\left(z,\tau\right)
 & = & 
 \ln\bar{w},
 \nonumber \\
 \Omega^{\mathrm{Kronecker}}_{2}\left(z,\tau\right)
 & = & 
 \ln\left(1-\bar{w}\right) - \frac{1}{2} \ln\bar{w} + \frac{1}{12} \ln\bar{q} - \bar{E}_{1,0}\left(\bar{w},1,\bar{q}\right),
 \nonumber \\
 \Omega^{\mathrm{Kronecker}}_{k}\left(z,\tau\right)
 & = & 
 \frac{B_{k-1}}{\left(k-1\right)!} \ln\bar{w} + \frac{\left(k-1\right)B_k}{k!} \ln\bar{q} - \frac{1}{\left(k-2\right)!}\bar{E}_{1,2-k}\left(\bar{w},1,\bar{q}\right),
 \;\;\; k \ge 3.
 \nonumber 
\eq


\section{The master integrals}
\label{sect:master_integrals}

\begin{figure}[!hbp]
\begin{center}
\includegraphics[scale=0.9]{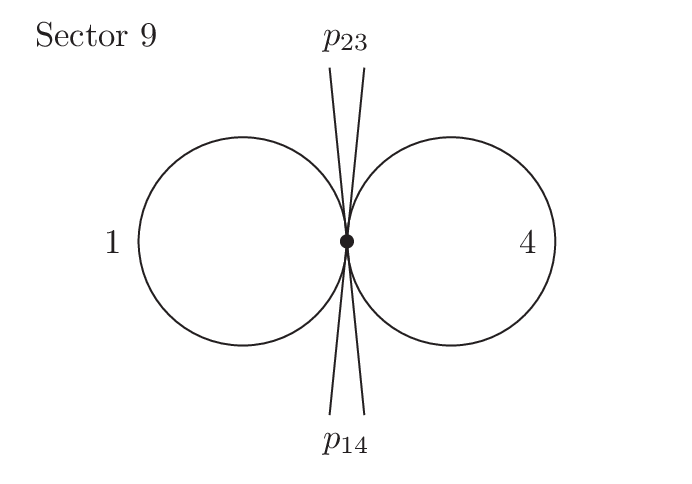}
\includegraphics[scale=0.9]{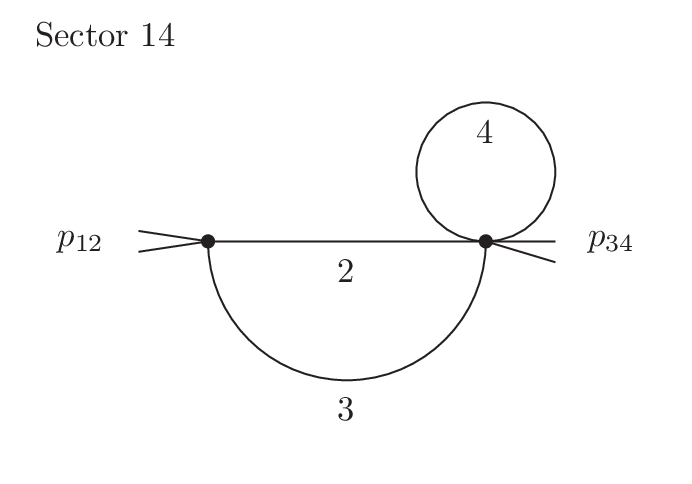}
\includegraphics[scale=0.9]{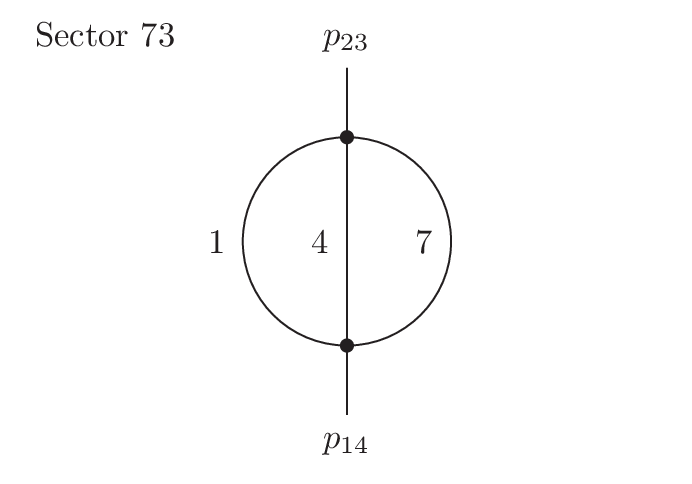}
\includegraphics[scale=0.9]{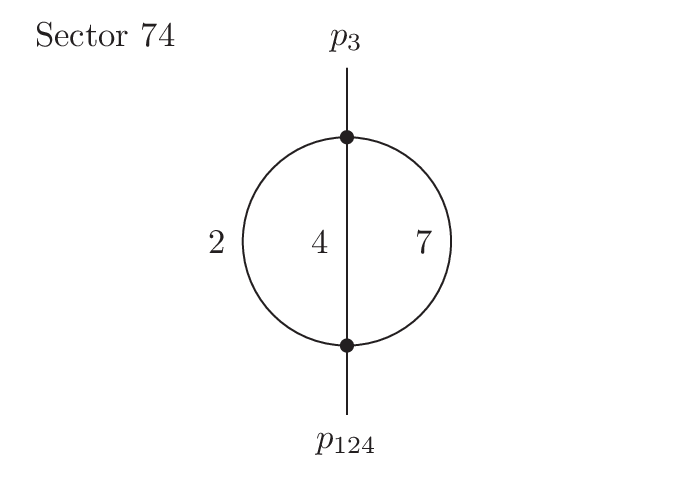}
\includegraphics[scale=0.9]{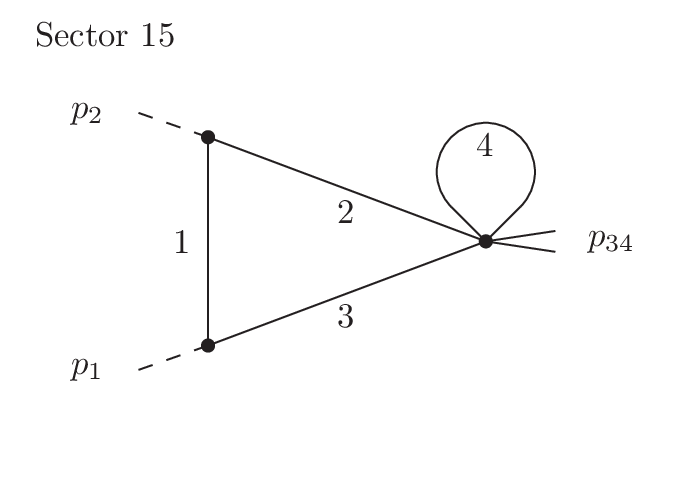}
\includegraphics[scale=0.9]{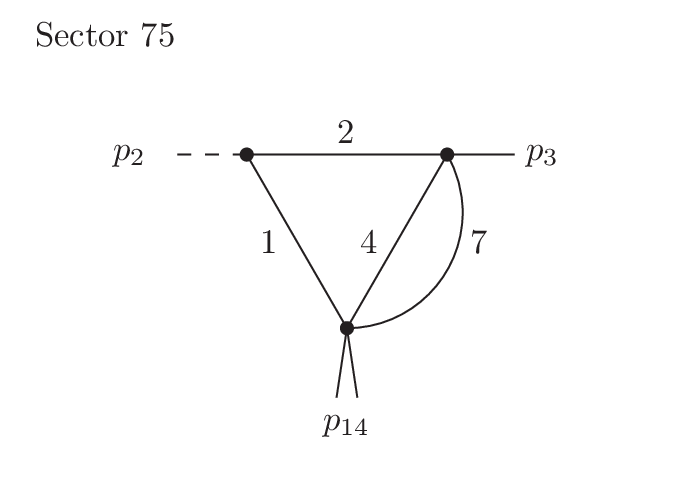}
\includegraphics[scale=0.9]{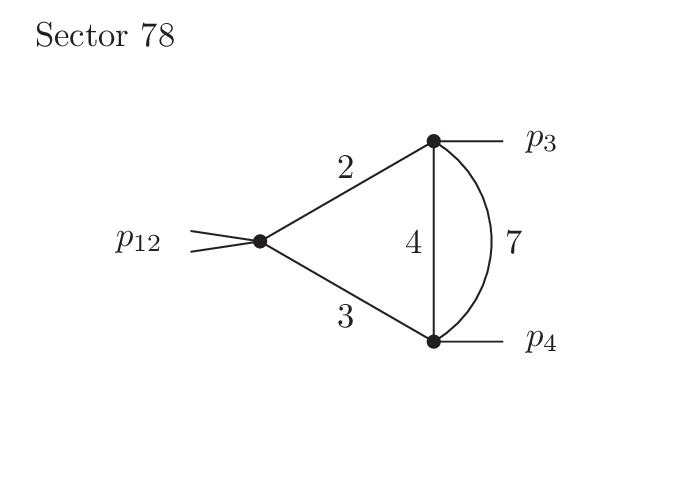}
\includegraphics[scale=0.9]{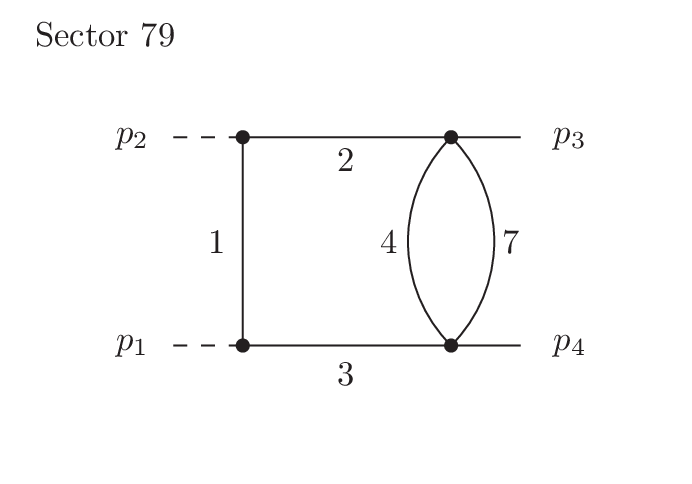}
\end{center}
\caption{
The diagrams for all master topologies.
In total there are $8$ master topologies.
A master topology may contain several master integrals.
This is the case for sector $73$ (two master integrals), sector $78$ (two master integrals)
and sector $79$ (three master integrals).
}
\label{fig_master_topologies}
\end{figure}

Using integration-by-parts identities \cite{Tkachov:1981wb,Chetyrkin:1981qh}
we may reduce any Feynman integral 
\bq
 I_{\nu_1 \nu_2 \nu_3 \nu_4 \nu_5 \nu_6 \nu_7 \nu_8 \nu_9}
\eq
(with $\nu_5, \nu_6, \nu_8, \nu_9 \le 0$)
to a linear combination of master integrals.
This reduction can be carried out with public available computer programs like
{\tt FIRE} \cite{Smirnov:2008iw,Smirnov:2019qkx},
{\tt Reduze} \cite{Studerus:2009ye,vonManteuffel:2012np} or
{\tt Kira} \cite{Maierhoefer:2017hyi,Klappert:2020nbg}.
There are $12$ master integrals for the sector $79$.
A straightforward basis of master integrals is for example
\bq
\label{def_non_canonical_basis}
 \vec{I}
 & = &
 \left(
 I_{100100000},
 I_{011100000},
 I_{100100100}, I_{200100100},
 I_{010100100}, 
 I_{111100000},
 I_{110100100},
 \right. \nonumber \\
 & & \left.
 I_{011100100}, I_{021100100}, 
 I_{111100100}, I_{211100100}, I_{121100100}
 \right)^T.
\eq
The corresponding diagrams are shown in fig.~\ref{fig_master_topologies}.

Differentiation under the integral sign combined with integration-by-parts identities allows us to derive
the differential equations for the master integrals.
For the basis $\vec{I}$ one finds
\bq
 d \vec{I} & = & \tilde{A} \vec{I},
\eq
where the $(12 \times 12)$-matrix $\tilde{A}$ depends on the kinematic variables and the dimensional regularisation
parameter $\eps$.
This differential equation is not in $\eps$-form.
The matrix $\tilde{A}$ is readily computable, albeit the result is not particularly aesthetic.

We seek another basis $\vec{J}$ of master integrals, such that the differential equation is of the form
\bq
 d \vec{J} & = & \eps A \vec{J},
\eq
where $A$ is now independent of the dimensional regularisation parameter $\eps$.
If the bases $\vec{J}$ and $\vec{I}$ are related by
\bq
 \vec{J} & = & U \vec{I},
\eq
the matrix $A$ is given by
\bq
 \eps A & = & U \tilde{A} U^{-1} - U d U^{-1}.
\eq
Given the existing literature for this family of Feynman integrals, it is not too difficult to find a basis $\vec{J}$ with this property.
In \cite{Adams:2018bsn,Adams:2018kez} a basis of master integrals is given, such that
\bq
 \tilde{A} & = & A^{(0)} + \eps A^{(1)},
\eq
such that $A^{(0)}$ and $A^{(1)}$ are independent of $\eps$ and in addition $A^{(0)}$ is strictly lower triangular.
In this case we find the sought-after basis $\vec{J}$ by integrating the non-zero entries of $A^{(0)}$.

\begin{figure}
\begin{center}
\includegraphics[scale=1.0]{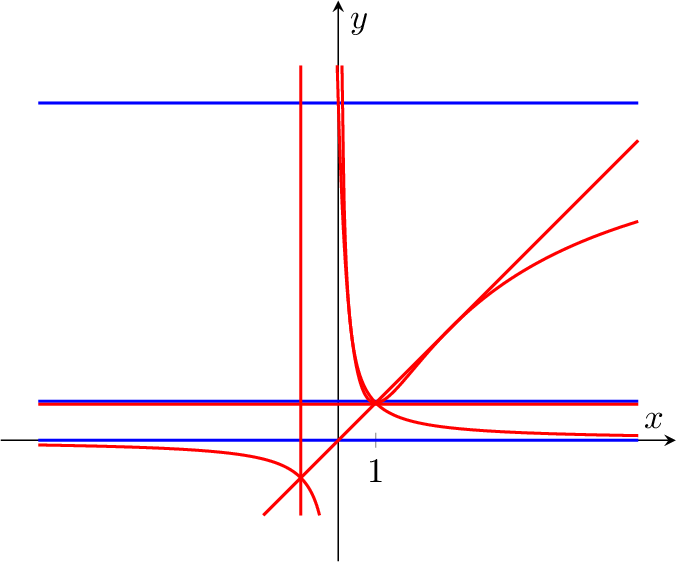}
\end{center}
\caption{
The location of the cusps in $(x,y)$-space for the curve $\curveone$ (shown in blue) and the curve $\curvetwo$ (shown in red).
}
\label{fig_sector_cusps}
\end{figure}

It will be convenient to perform in addition a minor modification 
by choosing a different pair of periods compared to \cite{Adams:2018bsn,Adams:2018kez}:
Compared to \cite{Adams:2018bsn,Adams:2018kez} we exchange the definitions of $Z_1$ and $Z_2$ (see eq.~(\ref{def_Z})). 
This amounts to a modular transformation
\bq
 \gamma & = & 
 \left(\begin{array}{rr}
  0 & -1 \\
  1 & 0 \\
 \end{array} \right).
\eq
The motivation for this modification is as follows:
We are going to analyse the system in a neighbourhood of $\bar{q}^{\curveone}=0$ and $\bar{q}^{\curvetwo}=0$.
For the choice of the periods as defined in section~\ref{sect:notation} the hyperplanes in $(x,y)$-space defined
by $\bar{q}^{\curveone}=0$ and $\bar{q}^{\curvetwo}=0$ intersect at $(x,y)=(0,0)$ with normal crossings, while
for the choice as in \cite{Adams:2018bsn,Adams:2018kez} the hyperplane defined by $\bar{q}^{\curveone}=0$ is contained
in a neighbourhood of $(x,y)=(-1,1)$ in the (reducible) hyperplane defined by $\bar{q}^{\curvetwo}=0$.
This can be inferred from the position of the cusps of the two elliptic curves in $(x,y)$-space.
For the curve $\curveone$ the cusps at finite distance are given by
\bq
 \mbox{Curve $\curveone$}:
 & &
 \left\{y=0\right\} \cup
 \left\{y=1\right\} \cup
 \left\{y=9\right\}.
\eq
For the curve $\curvetwo$ the cusps at finite distance are given by
\bq
 \mbox{Curve $\curvetwo$}:
 & &
 \left\{y=1\right\} \cup
 \left\{y=x\right\} \cup
 \left\{x+1=0\right\} \cup
 \left\{1-xy=0\right\} 
 \nonumber \\
 & &
 \cup
 \left\{x^2y-9x^2+2xy+14x+y-9=0\right\}.
\eq
The position of the cusps is shown in fig.~\ref{fig_sector_cusps}.

This minor modification leads again to a differential equation linear in $\eps$,
where the $\eps^0$-term is strictly lower triangular.
This minor modification is not essential, but it makes it easier to analyse the system of differential equations.

By integrating the $\eps^0$-terms we may transform the system to an $\eps$-form (\cite{Gehrmann:2014bfa}, see also chap. 7.1 of \cite{Weinzierl:2022eaz}).
We recall that ${\bf D}^-$ denotes the dimensional shift operator defined in eq.~(\ref{def_dimensional_shift}).
We arrive at 
\begin{alignat}{2}
 \mbox{Sector 9:} \;\;\;\; &
 J_{1}
 & = \;\; & 
 \eps^2 \; {\bf D}^- I_{100100000},
 \nonumber \\
 \mbox{Sector 14:} \;\;\;\; &
 J_{2}
 & = \;\; & 
 \eps^2 \frac{\left(1-x\right)\left(1+x\right)}{2x} \; {\bf D}^- I_{011100000},
 \nonumber \\
 \mbox{Sector 73:} \;\;\;\; &
 J_{3}
 & = \;\; & 
 \eps^2 \frac{\pi}{\psi^{\curveone}_1} \; {\bf D}^- I_{100100100},
 \nonumber \\
 &
 J_{4}
 & = \;\; & 
%
%
 \frac{\left(\psi^{\curveone}_1\right)^2}{2 \pi i \eps W^{\curveone}_y} \frac{d}{dy} J_{3} 
 + \frac{1}{24} \left(3y^2-10y-9\right) \left( \frac{\psi^{\curveone}_1}{\pi} \right)^2 J_{3},
 \nonumber \\
 \mbox{Sector 74:} \;\;\;\; &
 J_{5}
 & = \;\; & 
 4 \eps^2 \; {\bf D}^- I_{010100100},
 \nonumber \\
 \mbox{Sector 15:} \;\;\;\; &
 J_{6}
 & = \;\; & 
 - \eps^3 \left(1-\eps\right) \frac{\left(1-x\right)^2}{x} I_{111100000},
 \nonumber \\
 \mbox{Sector 75:} \;\;\;\; &
 J_{7}
 & = \;\; &
 \eps^3 \left(1-y\right) I_{110200100}, 
 \nonumber \\
 \mbox{Sector 78:} \;\;\;\; &
 J_{8}
 & = \;\; & 
 \eps^2 \frac{\left(1-x^2\right)^2}{x^2} I_{021200100}
 - \frac{3 \eps^2 \left(1-x\right)^2}{2 x} I_{020200100},
 \nonumber \\
 &
 J_{9}
 & = \;\; & 
 \eps^3 \frac{\left(1-x^2\right)}{x} I_{011200100},
 \nonumber \\
 \mbox{Sector 79:} \;\;\;\; &
 J_{10}
 & = \;\; & 
 \eps^3 \frac{\left(1-x\right)^2}{x} \frac{\pi}{\psi^{\curvetwo}_1} I_{111200100},
 \nonumber \\
 &
 J_{11}
 & = \;\; & 
 \eps^3 \left(1-2\eps\right) \frac{\left(1-x\right)^2}{x} I_{111100100}
 + F_{11,10} J_{10},
 \nonumber \\
 &
 J_{12}
 & = \;\; & 
 \frac{\left(\psi^{\curvetwo}_1\right)^2}{2 \pi i \eps W^{\curvetwo}_{y}} \frac{\partial}{\partial y} J_{10}
 + F_{12,11} \left( J_{11} - \frac{2}{3} J_{8} - \frac{4}{3} J_{7} - \frac{2}{3} J_{6} + \frac{1}{9} J_{5} \right)
 + F_{12,10} J_{10}
 \nonumber \\
 & & &
 + F_{12,3} J_{3}.
\end{alignat}
The functions $F_{11,10}$, $F_{12,11}$, $F_{12,10}$ and $F_{12,3}$ remove the $\eps^0$ terms and are obtained by integration.
We denote by $\gamma$ the integration path from $(x',y')=(1,0)$ first to $(1,y')$ (with $x'=1=\mathrm{const}$)
and then to $(x',y')$ along $y'=\mathrm{const}$.
\begin{figure}
\begin{center}
\includegraphics[scale=1.0]{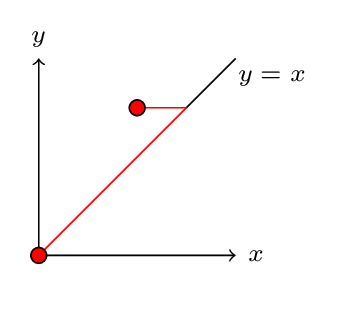}
\includegraphics[scale=1.0]{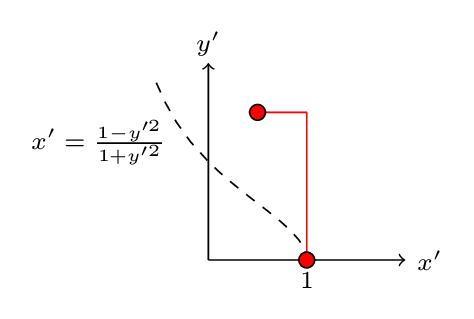}
\includegraphics[scale=1.0]{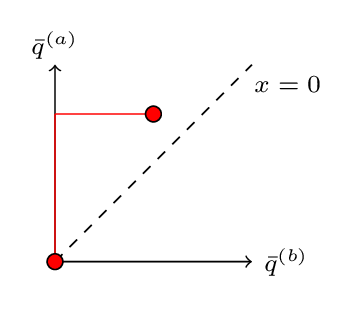}
\end{center}
\caption{
The integration path $\gamma$ (shown in red) in various coordinate systems.
}
\label{fig_integration_path}
\end{figure}
The integration path $\gamma$ is shown for various coordinate systems in fig.~\ref{fig_integration_path}.
The function $F_{11,10}$ is given by
\bq
\label{def_F_11_10}
 F_{11,10}
 & = & 
 \int\limits_{\gamma} f_{11,10},
 \nonumber \\
 f_{11,10}
 & = &
 - \frac{\psi_1^{(b)}}{\pi}
 \frac{1}{\left(3-4x-2xy+3x^2\right)} 
 \left[ 
       \frac{\left(1-y\right)\left(5-6x-y+5x^2-2xy-x^2y\right)}{\left(1-x^2\right)} dx 
       + \left(1-x\right)^2 dy 
 \right]
 \nonumber \\
 & &
 -
 \frac{\partial_y \psi_1^{\curvetwo}}{\pi}
 \frac{\left(9-14x-y-2xy+9x^2-x^2y\right)}{\left(1-x^2\right)\left(3-4x-2xy+3x^2\right)} 
 \left[ \left(1-y^2\right) dx - \left(1-x^2\right) dy \right].
\eq
The one-form $f_{11,10}$ is closed:
\bq
 d f_{11,10} & = & 0.
\eq
The functions $F_{12,11}$ and $F_{12,10}$ are related algebraically to $F_{11,10}$:
\bq
 F_{12,11}
 & = & 
 \frac{3}{8} \left[  - F_{11,10} + \frac{\psi_1^{(b)}}{\pi} \frac{\left(9-14x-y-2xy+9x^2-x^2y\right)}{\left(3-4x-2xy+3x^2\right)}\right],
 \nonumber \\
 F_{12,10}
 & = & 
 \frac{3}{16} \left[  - F_{11,10}^2 + \left( \frac{\psi_1^{(b)}}{\pi} \right)^2 \frac{P_{12,10}}{3 \left(1-x\right)^2 \left(3-4x-2xy+3x^2\right)}\right],
\eq
\bq
 P_{12,10}
 & = &
 63-238\,x-38\,y+358\,{x}^{2}+122\,yx+7\,{y}^{2}-238\,{x}^{3}-192\,y{x}^{2}-10\,{y}^{2}x+63\,{x}^{4}
 \nonumber \\
 & &
 +122\,y{x}^{3}+30\,{y}^{2}{x}^{2}-2\,{y}^{3}x-38\,y{x}^{4}-10\,{y}^{2}{x}^{3}-4\,{y}^{3}{x}^{2}+7\,{y}^{2}{x}^{4}-2\,{y}^{3}{x}^{3}.
\eq
The two algebraic relations follow from the observed relations 
\bq
 A_{10,10} \; = \; A_{12,12},
 & &
 8 A_{10,11} \; = \; 3 A_{11,12}.
\eq
With $z^{\curvetwo}$ defined in the next section, one finds that $F_{11,10}$ is given by
\bq
 F_{11,10}
 & = & 
 \frac{2}{\pi} g^{(1)}\left(z^{\curvetwo}+\frac{1}{6},\tau^{\curvetwo}\right).
\eq
We are free to choose an integration constant in eq.~(\ref{def_F_11_10}).
We require that $A_{10,10}$ (and $A_{12,12}$) reduce in the limit $x=0$ to $A_{3,3}$.
This fixes the integration constant to zero.
In this limit we then have (note that $\psi_1^{\curveone}=\psi_1^{\curvetwo}=\psi_1$ in this limit)
\bq
 \left. F_{11,10} \right|_{x=0} & = &
 - \frac{1}{3} \left(y-9\right) \frac{\psi_1}{\pi}.
\eq
The function $F_{12,3}$ is given by
\bq
\label{def_F_12_3}
 F_{12,3}
 & = &
 - \frac{1}{4}
 +
 \int\limits_{\gamma} f_{12,3},
 \nonumber \\
 f_{12,3}
 & = &
 \frac{1}{24}
 \frac{\psi_1^{\curveone}}{\pi} \frac{\psi_1^{\curvetwo}}{\pi}
 \frac{1}{\left(3-4x-2xy+3x^2\right)^2}
 \left[ 
  \frac{\left(1-x\right)\left(1-y\right)Q_1}{\left(1+x\right)} dx
  - Q_2 dy
 \right]
 \nonumber \\
 & &
 +
 \frac{1}{24}
 \frac{\partial_y \psi_1^{\curveone}}{\pi} \frac{\psi_1^{\curvetwo}}{\pi}
 \left[
   \frac{\left(1+x\right)y\left(1-y\right)\left(9-y\right)}{x\left(1-x\right)} dx
   - \frac{Q_3}{\left(3-4x-2xy+3x^2\right)} dy
 \right]
 \nonumber \\
 & & 
 +
 \frac{1}{24}
 \frac{\psi_1^{\curveone}}{\pi} \frac{\partial_y \psi_1^{\curvetwo}}{\pi}
   \frac{\left(1-y\right)\left(x-y\right)\left(1-xy\right)Q_4 Q_5}{x\left(1-x^2\right)\left(3-4x-2xy+3x^2\right)^2} dx
\eq
with
\bq
 Q_1
 & = &
 45-6\,x+30\,y+45\,{x}^{2}-52\,yx-3\,{y}^{2}+30\,y{x}^{2}-22\,{y}^{2}x-3\,{y}^{2}{x}^{2},
 \nonumber \\
 Q_2
 & = &
 45-108\,x-3\,y+142\,{x}^{2}-44\,yx-108\,{x}^{3}+62\,y{x}^{2}+45\,{x}^{4}-44\,y{x}^{3}+16\,{y}^{2}{x}^{2}-3\,y{x}^{4},
 \nonumber \\
 Q_3
 & = &
 27-42\,x+30\,y+27\,{x}^{2}-52\,yx-{y}^{2}+30\,y{x}^{2}-18\,{y}^{2}x-{y}^{2}{x}^{2},
 \nonumber \\
 Q_4
 & = & 
 9-6x-4xy+9x^2,
 \nonumber \\
 Q_5
 & = &
 9-14x-y+9x^2-2xy-x^2y.
\eq
The one-form $f_{12,3}$ is closed:
\bq
 d f_{12,3} & = & 0.
\eq
The integration constant for $F_{12,3}$ in eq.~(\ref{def_F_12_3}) 
has been chosen such that in the limit $x=0$ (corresponding to $\bar{q}^{\curveone}=\bar{q}^{\curvetwo}$)
we have
\bq
 \left. F_{12,3} \right|_{x=0} & = &
 \frac{1}{144} \left(y^2-30y-27\right) \frac{\psi_1^2}{\pi^2}.
\eq
As advertised, the differential equation for the basis $\vec{J}$ is in $\eps$-form:
\bq
\label{differential_equation_eps_form}
 d \vec{J} & = & \eps A \vec{J}.
\eq
The entries of the $(12 \times 12)$-matrix $A$ are differential one-forms.
The non-zero entries are
\bq
\lefteqn{
 A
 = } & \\
 & &
 \left(\begin{array}{cccccccccccc}
 0 & 0 & 0 & 0 & 0 & 0 & 0 & 0 & 0 & 0 & 0 & 0 \\
 A_{2,1} & A_{2,2} & 0 & 0 & 0 & 0 & 0 & 0 & 0 & 0 & 0 & 0 \\
 \cellcolor{yellow} 0 & \cellcolor{yellow} 0 & \cellcolor{yellow} A_{3,3} & \cellcolor{yellow} A_{3,4} & 0 & 0 & 0 & 0 & 0 & 0 & 0 & 0 \\
 \cellcolor{yellow} A_{4,1} & \cellcolor{yellow} 0 & \cellcolor{yellow} A_{4,3} & \cellcolor{yellow} A_{4,4} & 0 & 0 & 0 & 0 & 0 & 0 & 0 & 0 \\
 0 & 0 & \cellcolor{yellow} 0 & \cellcolor{yellow} 0 & 0 & 0 & 0 & 0 & 0 & 0 & 0 & 0 \\
 0 & A_{6,2} & \cellcolor{yellow} 0 & \cellcolor{yellow} 0 & 0 & 0 & 0 & 0 & 0 & 0 & 0 & 0 \\
 0 & 0 & \cellcolor{yellow} A_{7,3} & \cellcolor{yellow} 0 & 0 & 0 & 0 & 0 & 0 & 0 & 0 & 0 \\
 0 & A_{8,2} & \cellcolor{yellow} 0 & \cellcolor{yellow} 0 & A_{8,5} & 0 & 0 & A_{8,8} & A_{8,9} & 0 & 0 & 0 \\
 0 & 0 & \cellcolor{yellow} 0 & \cellcolor{yellow} 0 & A_{9,5} & 0 & 0 & A_{9,8} & A_{9,9} & 0 & 0 & 0 \\
 \cellcolor{red} 0 & \cellcolor{red} 0 & \cellcolor{orange} A_{10,3} & \cellcolor{orange} 0 & \cellcolor{red} A_{10,5} & \cellcolor{red} A_{10,6} & \cellcolor{red} A_{10,7} & \cellcolor{red} A_{10,8} & \cellcolor{red} 0 & \cellcolor{red} A_{10,10} & \cellcolor{red} A_{10,11} & \cellcolor{red} A_{10,12} \\
 \cellcolor{red} 0 & \cellcolor{red} A_{11,2} & \cellcolor{orange} A_{11,3} & \cellcolor{orange} 0 & \cellcolor{red} A_{11,5} & \cellcolor{red} A_{11,6} & \cellcolor{red} A_{11,7} & \cellcolor{red} A_{11,8} & \cellcolor{red} A_{11,9} & \cellcolor{red} A_{11,10} & \cellcolor{red} A_{11,11} & \cellcolor{red} A_{11,12} \\
 \cellcolor{red} 0 & \cellcolor{red} A_{12,2} & \cellcolor{orange} A_{12,3} & \cellcolor{orange} A_{12,4} & \cellcolor{red} A_{12,5} & \cellcolor{red} A_{12,6} & \cellcolor{red} A_{12,7} & \cellcolor{red} A_{12,8} & \cellcolor{red} A_{12,9} & \cellcolor{red} A_{12,10} & \cellcolor{red} A_{12,11} & \cellcolor{red} A_{12,12} \\
 \end{array} \right).
 \nonumber
\eq
The colour coding is as follows:
Entries not highlighted by any colour are dlog-forms.
Entries highlighted in yellow are related to curve $\curveone$, but independent of curve $\curvetwo$.
These entries are of the form 
\bq
 f_k(\tau^{\curveone}) \; 2\pi i d\tau^{\curveone},
\eq
where $f_k(\tau^{\curveone})$ is a modular form for curve $\curveone$.
Entries highlighted in red are related to curve $\curvetwo$, but independent of curve $\curveone$.
These are of the form
\bq
 f_k(\tau^{\curvetwo}) \; 2\pi i d\tau^{\curvetwo},
 & &
 \omega^{\mathrm{Kronecker}}_{k}\left(az^{\curvetwo}+b,\tau^{\curvetwo}\right),
\eq
with $\omega^{\mathrm{Kronecker}}_{k}$ defined in eq.~(\ref{def_omega_Kronecker}).

The entries highlighted in orange depend on both elliptic curves and are the most interesting ones.
The non-zero entries are 
$A_{10,3}$, $A_{11,3}$, $A_{12,3}$ and $A_{12,4}$.
We call them the mixed terms and all other entries the non-mixed terms.


\section{The non-mixed entries of the differential equation}
\label{sect:non-mixed}

Let us first discuss the non-mixed terms appearing in the differential equation.
We start with the dlog-forms, followed by the terms related to curve $\curveone$,
and finish with the terms related to curve $\curvetwo$.

The entries of the matrix $A$ in the lines $1-9$ are all well-known.
They depend either only on $s$ or only on $t$, but not on both variables.
The entries which only depend on $s$ are all dlog-forms.
The entries which only depend on $t$ are proportional to modular forms of $\Gamma_1(6)$.
The non-mixed entries in the lines $10-12$ depend on both variables $s$ and $t$.
We may express them in terms of differential one-forms related to the moduli space ${\mathcal M}_{1,2}$.
Coordinates on this moduli space are $(\tau^{\curvetwo},z^{\curvetwo})$.
We construct $z^{\curvetwo}$ in section~\ref{sect:constructing_z}.

\subsection{The dlog-forms}

We start with the dlog-forms.
We introduce
\begin{alignat}{3}
\label{def_omega}
 &
 \omega^{\mathrm{mpl}}_{s,0}
 & \; = \; &
 \frac{ds}{s} 
 & \; = \; & 
 \frac{2 dx}{x-1} - \frac{dx}{x},
 \nonumber \\
 &
 \omega^{\mathrm{mpl}}_{s,4}
 & \; = \; &
 \frac{ds}{s-4m^2}
 & \; = \; & 
 \frac{2 dx}{x+1} - \frac{dx}{x},
 \nonumber \\
 &
 \omega^{\mathrm{mpl}}_{s,0,4}
 & \; = \; &
 \frac{ds}{\sqrt{-s\left(4m^2-s\right)}}
 & \; = \; &
 \frac{dx}{x}.
\end{alignat}
Then
\begin{align}
 A_{2,1} & = - \omega^{\mathrm{mpl}}_{s,0,4},
 &
 A_{2,2} & = - \omega^{\mathrm{mpl}}_{s,4},
 \nonumber \\
 A_{6,2} & = - \omega^{\mathrm{mpl}}_{s,0,4},
 \nonumber \\
 A_{8,2} & = \omega^{\mathrm{mpl}}_{s,0,4},
 &
 A_{8,5} & = \frac{1}{2} \omega^{\mathrm{mpl}}_{s,4},
 &
 A_{8,8} & = -2 \omega^{\mathrm{mpl}}_{s,0} - \omega^{\mathrm{mpl}}_{s,4},
 &
 A_{8,9} & = - 3 \omega^{\mathrm{mpl}}_{s,0,4},
 \nonumber \\
 A_{9,5} & = - \frac{1}{6} \omega^{\mathrm{mpl}}_{s,0,4},
 &
 A_{9,8} & = \omega^{\mathrm{mpl}}_{s,0,4},
 &
 A_{9,9} & = \omega^{\mathrm{mpl}}_{s,4}.
\end{align}

\subsection{The entries related to curve $\curveone$}

We introduce
\begin{alignat}{4}
\label{def_modular}
 &
 \omega_0^{\mathrm{modular}, \curveone}
 & \; = \; &
 & & 
 2 \pi i \; d\tau^{\curveone}
 & \; = \; & 
 \frac{2\pi i W_y^{\curveone}}{\left(\psi_1^{\curveone}\right)^2} dy,
 \nonumber \\
 &
 \omega^{\mathrm{modular}, \curveone}_{2,0}
 & \; = \; &
 \omega^{\mathrm{mpl}}_{t,0}
 & \; = \; & 
 g_{2,0}(\tau^{\curveone}) \; \left(2\pi i\right) d\tau^{\curveone}
 & \; = \; & 
 \frac{dy}{y},
 \nonumber \\
 &
 \omega^{\mathrm{modular}, \curveone}_{2,1}
 & \; = \; &
 \omega^{\mathrm{mpl}}_{t,1}
 & \; = \; & 
 g_{2,1}(\tau^{\curveone}) \; \left(2\pi i\right) d\tau^{\curveone}
 & \; = \; & 
 \frac{dy}{y-1},
 \nonumber \\
 &
 \omega^{\mathrm{modular}, \curveone}_{2,9}
 & \; = \; &
 \omega^{\mathrm{mpl}}_{t,9}
 & \; = \; & 
 g_{2,9}(\tau^{\curveone}) \; \left(2\pi i\right) d\tau^{\curveone}
 & \; = \; & 
 \frac{dy}{y-9},
 \nonumber \\
 &
 \omega^{\mathrm{modular}, \curveone}_{3}
 & \; = \; &
 & & 
 g_{3}(\tau^{\curveone}) \; \left(2\pi i\right) d\tau^{\curveone}
 & \; = \; & 
 \frac{\psi_1^{\curveone}}{\pi} dy,
 \nonumber \\
 &
 \omega^{\mathrm{modular}, \curveone}_{4}
 & \; = \; &
 & & 
 f_{4}(\tau^{\curveone}) \; \left(2\pi i\right) d\tau^{\curveone}
 & \; = \; & 
 \frac{\left(y+3\right)^4}{48 y \left(y-1\right) \left(y-9\right)} \left( \frac{\psi_1^{\curveone}}{\pi} \right)^2 dy.
\end{alignat}
$g_{2,0}$, $g_{2,1}$, $g_{2,9}$, $g_{3}$ and $f_{4}$ are modular forms of $\Gamma_1(6)$.
$\omega^{\mathrm{modular}, \curveone}_{2,0}$, $\omega^{\mathrm{modular}, \curveone}_{2,1}$ and $\omega^{\mathrm{modular}, \curveone}_{2,9}$ are also dlog-forms
and we denote them alternatively by $\omega^{\mathrm{mpl}}_{t,0}$, $\omega^{\mathrm{mpl}}_{t,1}$ and $\omega^{\mathrm{mpl}}_{t,9}$, respectively.
We have
\begin{align}
 A_{3,4} & = \omega_0^{\mathrm{modular}, \curveone},
 \nonumber \\
 A_{3,3} & = \frac{1}{2} \omega^{\mathrm{mpl}}_{t,0} - \omega^{\mathrm{mpl}}_{t,1} - \omega^{\mathrm{mpl}}_{t,9},
 &
 A_{4.4} & = \frac{1}{2} \omega^{\mathrm{mpl}}_{t,0} - \omega^{\mathrm{mpl}}_{t,1} - \omega^{\mathrm{mpl}}_{t,9},
 \nonumber \\
 A_{4,1} & = - \frac{1}{2} \omega^{\mathrm{modular}, \curveone}_{3},
 &
 A_{7,3} & = - \frac{1}{3} \omega^{\mathrm{modular}, \curveone}_{3},
 \nonumber \\
 A_{4,3} & = \omega^{\mathrm{modular}, \curveone}_{4}.
\end{align}

\subsection{The entries related to curve $\curvetwo$}
\label{sect:non-mixed_curve_b}

Let us now turn to the non-mixed entries in lines $10-12$.
This concerns the columns $2$ and $5-12$.
In order to express these entries we introduce differential one-forms, which can be grouped into three
categories: dlog-forms, differential one-form proportional to modular forms (for curve $\curvetwo$)
and differential one-forms related to the coefficients of the Kronecker function. 
For the latter we first define an additional variable $z^{\curvetwo}$.
Together with $\tau^{\curvetwo}$ the pair $(\tau^{\curvetwo},z^{\curvetwo})$ defines the standard coordinates
on the moduli space ${\mathcal M}_{1,2}$ of a genus one curve with two marked points.
The variable $\tau^{\curvetwo}$ parametrises the shape of the curve, the variable $z^{\curvetwo}$ gives the location of one marked point on the curve.
By translation invariance, the other marked point can be fixed at the origin.

\subsubsection{Constructing $z^{\curvetwo}$}
\label{sect:constructing_z}

In this paragraph we construct $z^{\curvetwo}$.
By modular weight counting we expect $A_{10,11}$ to be of modular weight $1$ with respect to curve $\curvetwo$.
On ${\mathcal M}_{1,2}$ the only differential one-form of modular weight $1$ is
\bq
 \omega_1(z^{\curvetwo},\tau^{\curvetwo}) & = & 2\pi i dz^{\curvetwo}.
\eq
We therefore expect $A_{10,11}$ to be proportional to $\omega_1(z^{\curvetwo},\tau^{\curvetwo})$:
\bq
 dz^{\curvetwo}
 & = & 
 \frac{\lambda}{2 \pi i} A_{10,11},
\eq
where $\lambda$ denotes the constant of proportionality.
Integration gives
\bq
 z^{\curvetwo}
 & = &
 z_0^{\curvetwo}
 + \frac{\lambda}{2 \pi i} \left[
    \frac{3}{2i}\ln\left(\frac{\sqrt{\left(1+y'^2\right)\left(3-y'^2\right)}-i\left(1-y'^2\right)}{2}\right)
 \right. \\
 & & \left.
    - \frac{9}{2} \left(1-y'^2\right) \sqrt{\left(1+y'^2\right)\left(3-y'^2\right)} \bar{q}^{\curvetwo}
 \right. \nonumber \\
 & & \left.
    - \frac{3}{4} \left(1-y'^2\right) \left(3+70y'^2-35y'^4\right) \sqrt{\left(1+y'^2\right)\left(3-y'^2\right)} \left( \bar{q}^{\curvetwo} \right)^2
   \right]
 + {\mathcal O}\left( \left( \bar{q}^{\curvetwo} \right)^3 \right).
 \nonumber 
\eq
$z_0^{\curvetwo}$ and $\lambda$ are (at the moment) two unknown constants.
We make the ad-hoc choice
\bq
\label{value_z_0_lambda}
 z_0^{\curvetwo} \; = \; \frac{1}{12},
 & &
 \lambda \; = \; \frac{2}{3} i.
\eq
This choice is motivated by the following two observations:
It is easily checked that $dz^{\curvetwo}=0$ on the hypersurface $x'=(1-y'^2)/(1+y'^2)$ and 
therefore $z^{\curvetwo}$ is constant there. 
We require that on the hypersurface $x'=(1-y'^2)/(1+y'^2)$ we have
\bq
 z^{\curvetwo} & = & 0.
\eq
This gives the relation
\bq
 z_0^{\curvetwo} & = & - \frac{1}{8} i \lambda.
\eq
We obtain additional information on $\lambda$ as follows:
We define
\bq
 \bar{w}^{\curvetwo} & = & e^{2 \pi i z^{\curvetwo}}.
\eq
Then
\bq
 \bar{w}^{\curvetwo}
 & = &
 e^{2\pi i z_0^{\curvetwo}} 
 \left[ \left(\frac{\sqrt{\left(1+y'^2\right)\left(3-y'^2\right)}-i\left(1-y'^2\right)}{2}\right)^{\frac{3\lambda}{2i}} 
 + {\mathcal O}\left(\bar{q}^{\curvetwo}\right) \right].
\eq
We expect the exponent $\frac{3\lambda}{2i}$ to be a rational number.
The simplest choice would be 
that the exponent is $\pm 1$.
Requiring that the exponent equals $1$ gives the value of $\lambda$ given in eq.~(\ref{value_z_0_lambda}).
In summary, we define $z^{\curvetwo}$ by
\bq
 z^{\curvetwo}
 & = &
 \frac{1}{12}
 + \frac{1}{3 \pi} \int\limits_\gamma A_{10,11}.
\eq
The first few terms in the expansion in $\bar{q}^{\curveone}$ and $\bar{q}^{\curvetwo}$ are
\bq
 z^{\curvetwo}
 & = &
 \frac{3\sqrt{3}}{2\pi}
 \left\{
        \bar{q}^{\curveone} 
        -\bar{q}^{\curvetwo} 
        - \frac{11}{2} \left( \bar{q}^{\curveone} \right)^2
        + 6 \bar{q}^{\curveone} \bar{q}^{\curvetwo}
        - \frac{1}{2} \left( \bar{q}^{\curvetwo} \right)^2
        + 31 \left( \bar{q}^{\curveone} \right)^3
 \right. \nonumber \\
 & & \left.
        + 21 \left( \bar{q}^{\curveone} \right)^2 \bar{q}^{\curvetwo} 
        - 102 \bar{q}^{\curveone} \left( \bar{q}^{\curvetwo} \right)^2
        + 50 \left( \bar{q}^{\curvetwo} \right)^3
        + \dots
 \right\}.
\eq
We remark that the concrete choice of $z_0$ and $\lambda$ does not really matter, we just have to make some choice.
We will soon see that the $z$-arguments of the Kronecker functions $g^{(k)}(z,\tau)$ are
linear functions
\bq
 \alpha z^{\curvetwo} + \beta.
\eq
The constants $\alpha$ and $\beta$ will be different for different choices of $z_0$ and $\lambda$
and compensate any ad-hoc choice of $z_0$ and $\lambda$.


\subsubsection{The expressions for the entries related to curve $\curvetwo$}

Having defined $z^{\curvetwo}$ we may now give the expressions for the non-mixed entries in the lines $10-12$ of the matrix $A$.
In order to express these entries we introduce differential one-forms, which can be grouped into three
categories: dlog-forms, differential one-form proportional to modular forms
and differential one-forms related to the coefficients of the Kronecker function. 
We start with the simplest differential one-forms. These are the dlog-forms. 
In addition to the dlog-forms already introduced we will need dlog-forms, which depend on both variables
$s$ and $t$.
We introduce
\bq
 \omega^{\mathrm{mpl}}_{s,t,1}
 & = &
 d \ln\left(x-y\right) + d \ln\left(xy-1\right),
 \nonumber \\
 \omega^{\mathrm{mpl}}_{s,t,2}
 & = &
 d \ln\left(xy-1\right).
\eq
In addition, we will encounter differential one-forms, which are proportional to modular forms.
These depend only on a single variable $\tau^{\curvetwo}$.
They are given by
\begin{alignat}{3}
\label{def_modular_b}
 &
 \omega_0^{\mathrm{modular}, \curvetwo}
 & \; = \; &
 2 \pi i \; d\tau^{\curvetwo}
 & \; = \; & 
 \frac{2\pi i}{\left(\psi_1^{\curvetwo}\right)^2} \left( W_x^{\curvetwo} dx + W_y^{\curvetwo} dy \right),
 \nonumber \\
 &
 \omega^{\mathrm{modular}, \curvetwo}_{2}
 & \; = \; &
 b_2(\tau^{\curvetwo}) \; \frac{d\tau^{\curvetwo}}{2\pi i},
 & & 
 \nonumber \\
 &
 \omega^{\mathrm{modular}, \curvetwo}_{4}
 & \; = \; &
 e_4(\tau^{\curvetwo}) \; \frac{d\tau^{\curvetwo}}{\left(2\pi i\right)^3}.
 & & 
\end{alignat}
$e_k(\tau)$ denotes the standard Eisenstein series
\bq
 e_k\left(\tau\right)
 & = &
 \sideset{}{_e}\sum\limits_{(n_1,n_2) \in {\mathbb Z}^2\backslash (0,0)} \frac{1}{\left(n_1 + n_2\tau \right)^k},
\eq
where the subscript $e$ at the summation symbol indicates that Eisenstein's summation prescription is understood.
$b_2(\tau)$ is defined by
\bq
 b_2\left(\tau\right)
 & = &
 e_2\left(\tau\right) - 2  e_2\left(2\tau\right).
\eq
$b_2(\tau)$ is a modular form of $\Gamma_0(2)$, $e_4(\tau)$ is a modular form of $\mathrm{SL}_2({\mathbb Z})$.

For the differential one-forms related to the coefficients of the Kronecker function
we note that at weight zero and one we have
\bq
 \omega^{\mathrm{Kronecker}}_{0}\left( z^{\curvetwo}, \tau^{\curvetwo} \right)
 & = &
 - \omega_0^{\mathrm{modular}, \curvetwo}
 \; = \; 
 - 2 \pi i \; d\tau^{\curvetwo},
 \nonumber \\
 \omega^{\mathrm{Kronecker}}_{1}\left( z^{\curvetwo}, \tau^{\curvetwo} \right)
 & = &
 2 \pi i \; dz^{\curvetwo}.
\eq
At modular weight $0$ with respect to curve $\curvetwo$ we have
\bq
 A_{10,12} 
 & = & 
 \omega_0^{\mathrm{modular}, \curvetwo}.
\eq
At modular weight $1$ with respect to curve $\curvetwo$ we have
\begin{align}
 A_{10,5} & = -\frac{i}{6} \omega^{\mathrm{Kronecker}}_{1}\left(z^{\curvetwo},\tau^{\curvetwo}\right),
 &
 A_{10,6} & = i \omega^{\mathrm{Kronecker}}_{1}\left(z^{\curvetwo},\tau^{\curvetwo}\right),
 \nonumber \\
 A_{10,7} & = 2 i \omega^{\mathrm{Kronecker}}_{1}\left(z^{\curvetwo},\tau^{\curvetwo}\right),
 &
 A_{10,8} 
 & =
 i \omega^{\mathrm{Kronecker}}_{1}\left(z^{\curvetwo},\tau^{\curvetwo}\right),
 \nonumber \\
 A_{10,11} 
 & = 
 - \frac{3i}{2} \omega^{\mathrm{Kronecker}}_{1}\left(z^{\curvetwo},\tau^{\curvetwo}\right),
 &
 A_{11,12} 
 & = 
 - 4 i \omega^{\mathrm{Kronecker}}_{1}\left(z^{\curvetwo},\tau^{\curvetwo}\right).
\end{align}
At modular weight $2$ with respect to curve $\curvetwo$ we have
\bq
 A_{10,10} 
 & = &
 - 6 \omega^{\mathrm{Kronecker}}_{2}\left(z^{\curvetwo}+\frac{2}{3},\tau^{\curvetwo}\right)
 + \omega^{\mathrm{mpl}}_{s,t,1}
 - 2 \omega^{\mathrm{mpl}}_{t,1}
 - \omega^{\mathrm{mpl}}_{s,4}
 - \omega^{\mathrm{mpl}}_{s,0,4},
 \nonumber \\\\
 A_{11,2} 
 & = & 
 - \omega^{\mathrm{mpl}}_{s,0,4},
 \nonumber \\
 A_{11,5} 
 & = &
 \frac{4}{3} \omega^{\mathrm{Kronecker}}_{2}\left(z^{\curvetwo}+\frac{2}{3},\tau^{\curvetwo}\right)
 + \frac{8}{3} \omega^{\mathrm{modular}, \curvetwo}_{2}
 - \frac{1}{3} \omega^{\mathrm{mpl}}_{s,t,1}
 + \omega^{\mathrm{mpl}}_{t,1}
 - \frac{1}{6} \omega^{\mathrm{mpl}}_{s,0}
 + \frac{1}{3} \omega^{\mathrm{mpl}}_{s,4}
 \nonumber \\
 & &
 + \frac{1}{3} \omega^{\mathrm{mpl}}_{s,0,4},
 \nonumber \\
 A_{11,6} 
 & = &
 - 8 \omega^{\mathrm{Kronecker}}_{2}\left(z^{\curvetwo}+\frac{2}{3},\tau^{\curvetwo}\right)
 - 16 \omega^{\mathrm{modular}, \curvetwo}_{2}
 + \omega^{\mathrm{mpl}}_{s,t,1}
 - 4 \omega^{\mathrm{mpl}}_{t,1}
 + \omega^{\mathrm{mpl}}_{s,0}
 - 2 \omega^{\mathrm{mpl}}_{s,4}
 \nonumber \\
 & &
 - \omega^{\mathrm{mpl}}_{s,0,4},
 \nonumber \\
 A_{11,9} 
 & = &
 \omega^{\mathrm{mpl}}_{s,t,1}
 - 2 \omega^{\mathrm{mpl}}_{s,t,2}
 + \omega^{\mathrm{mpl}}_{s,0,4},
 \nonumber \\
 A_{11,11} 
 & = &
 12 \omega^{\mathrm{Kronecker}}_{2}\left(z^{\curvetwo}+\frac{2}{3},\tau^{\curvetwo}\right)
 + 24 \omega^{\mathrm{modular}, \curvetwo}_{2}
 - 2 \omega^{\mathrm{mpl}}_{s,t,1}
 + 6 \omega^{\mathrm{mpl}}_{t,1}
 - \omega^{\mathrm{mpl}}_{s,0}
 + 3 \omega^{\mathrm{mpl}}_{s,4}
 \nonumber \\
 & &
 + 2 \omega^{\mathrm{mpl}}_{s,0,4}
\eq
and
\bq
 A_{12,12}  \; = \; A_{10,10},
 \;\;\;\;\;\;
 A_{11,7}  \; = \; 2 A_{11,6},
 \;\;\;\;\;\;
 A_{11,8}  \; = \; -6 A_{11,5}.
\eq
All dlog-forms appearing at modular weight $2$ can be expressed as a linear combination
of forms 
\bq
 \omega^{\mathrm{Kronecker}}_{2}(az^{\curvetwo}+b,\tau^{\curvetwo}).
\eq
The relevant formulae are given in appendix~\ref{sect:dlog_forms}.

At modular weight $3$ with respect to curve $\curvetwo$ we have
\bq
 A_{11,10}
 & = &
  - 8 i\omega^{\mathrm{Kronecker}}_{3}\left(2z^{\curvetwo}+\frac{1}{3},\tau^{\curvetwo}\right)
  - 8 i \omega^{\mathrm{Kronecker}}_{3}\left(z^{\curvetwo}+\frac{2}{3},\tau^{\curvetwo}\right),
 \nonumber \\
 A_{12,2} 
 & = & 
  \frac{i}{2} \omega^{\mathrm{Kronecker},\frac{1}{2}}_{3}\left(3z^{\curvetwo},\tau^{\curvetwo}\right)
  - \frac{3i}{2} \omega^{\mathrm{Kronecker},\frac{1}{2}}_{3}\left(z^{\curvetwo}+\frac{2}{3},\tau^{\curvetwo}\right),
 \nonumber \\
 A_{12,5}
 & = &
  - \frac{i}{4} \omega^{\mathrm{Kronecker}}_{3}\left(3z^{\curvetwo},\tau^{\curvetwo}\right)
  + \frac{i}{3} \omega^{\mathrm{Kronecker}}_{3}\left(2z^{\curvetwo}+\frac{1}{3},\tau^{\curvetwo}\right)
 \nonumber \\
 & &
  - \frac{i}{4} \omega^{\mathrm{Kronecker}}_{3}\left(z^{\curvetwo}+\frac{2}{3},\tau^{\curvetwo}\right)
  + \frac{4i}{3} \omega^{\mathrm{Kronecker}}_{3}\left(z^{\curvetwo}+\frac{1}{6},\tau^{\curvetwo}\right),
 \nonumber \\
 A_{12,6}
 & = &
  i \omega^{\mathrm{Kronecker}}_{3}\left(2z^{\curvetwo}+\frac{1}{3},\tau^{\curvetwo}\right)
  + 6 i \omega^{\mathrm{Kronecker}}_{3}\left(z^{\curvetwo}+\frac{2}{3},\tau^{\curvetwo}\right)
 \nonumber \\
 & &
  + 4 i \omega^{\mathrm{Kronecker}}_{3}\left(z^{\curvetwo}+\frac{1}{6},\tau^{\curvetwo}\right),
 \nonumber \\
 A_{12,8}
 & = &
  \frac{3i}{2} \omega^{\mathrm{Kronecker}}_{3}\left(3z^{\curvetwo},\tau^{\curvetwo}\right)
  - 2 i          \omega^{\mathrm{Kronecker}}_{3}\left(2z^{\curvetwo}+\frac{5}{6},\tau^{\curvetwo}\right)
 \nonumber \\
 & &
  - \frac{5i}{2} \omega^{\mathrm{Kronecker}}_{3}\left(z^{\curvetwo}+\frac{2}{3},\tau^{\curvetwo}\right)
  - 4 i          \omega^{\mathrm{Kronecker}}_{3}\left(z^{\curvetwo}+\frac{1}{6},\tau^{\curvetwo}\right),
 \nonumber \\
 A_{12,9}
 & = &
  - \frac{3i}{2} \omega^{\mathrm{Kronecker},\frac{1}{2}}_{3}\left(3z^{\curvetwo},\tau^{\curvetwo}\right)
  - \frac{3i}{2} \omega^{\mathrm{Kronecker},\frac{1}{2}}_{3}\left(z^{\curvetwo}+\frac{2}{3},\tau^{\curvetwo}\right)
\eq
and
\bq
 A_{12,11}  \; = \;  \frac{3}{8} A_{11,10},
 \;\;\;\;\;\;
 A_{12,7} \; = \; 2  A_{12,6}.
\eq
At modular weight $4$ with respect to curve $\curvetwo$ we have
\bq
 A_{12,10}
 & = &
  - 12 \omega^{\mathrm{Kronecker}}_{4}\left(2z^{\curvetwo}+\frac{1}{3},\tau^{\curvetwo}\right)
  - 24 \omega^{\mathrm{Kronecker}}_{4}\left(z^{\curvetwo}+\frac{2}{3},\tau^{\curvetwo}\right)
  + 72 \omega^{\mathrm{modular}, \curvetwo}_{4}.
 \nonumber \\
\eq


\section{The mixed entries of the differential equation}
\label{sect:mixed}

In this section we discuss the mixed entries
\bq
 A_{10,3}, 
 \;\;\;
 A_{11,3}, 
 \;\;\;
 A_{12,3}, 
 \;\;\;
 A_{12,4}.
\eq
These depend on both elliptic curves $\curveone$ and $\curvetwo$.

We study various representations of these entries.
The most straightforward representation is the one in the variables $(x,y)$. 
This representation is directly obtained from the differential equation.
The representation of the mixed entries in the $(x,y)$-coordinates is given in appendix~\ref{sect:mixed_xy}.

We also would like to express the mixed entries in the coordinates $\tau^{\curveone}$, $\tau^{\curvetwo}$ and $z^{\curvetwo}$.
These coordinates are more natural from a mathematical point of view.
As our Feynman integral depends only on two kinematic variables, the three variables
$\tau^{\curveone}, \tau^{\curvetwo}, z^{\curvetwo}$
are not independent.
We may choose $\tau^{\curveone}$ and $\tau^{\curvetwo}$ as our basic variables and express $z^{\curvetwo}$
as a function of the first two:
\bq
 z^{\curvetwo} & = & z^{\curvetwo}\left(\tau^{\curveone}, \tau^{\curvetwo}\right).
\eq
In this way we obtain a double series expansion in $\bar{q}^{\curveone}$ and $\bar{q}^{\curvetwo}$.
The result is given in section~\ref{sect:expansion}.
This result is particular useful for numerical evaluations.

On the more formal side we also would like to have a representation, which makes the modular transformation properties
under modular transformations with respect to curve $\curveone$ and independently curve $\curvetwo$ more transparent.
Thus we seek a representation in terms of the three variables $\tau^{\curveone}, \tau^{\curvetwo}, z^{\curvetwo}$,
with the understanding that one variable is a function of the two others.
Such a representation is of course not unique, as we only require that these representations
agree on a two-dimensional hypersurface inside a three-dimensional space.
It turns out, that integrability together with constraints from degenerate limits gives a natural
representation in terms of the three variables $\tau^{\curveone}, \tau^{\curvetwo}, z^{\curvetwo}$.
We study degenerate limits in section~\ref{sect:limits}
and integrability in section~\ref{sect:integrabilty}.

\subsection{Expansions}
\label{sect:expansion}

We start with the practical aspects: For numerical evaluations we would like to know 
the expansions of the differential one-forms $A_{10,3}$, $A_{11,3}$, $A_{12,3}$ and $A_{12,4}$ in 
$\bar{q}^{\curveone}$ and $\bar{q}^{\curvetwo}$.
The first few terms can be obtained without any problems.
In order to present the mixed entries in a compact form, we introduce primitives by
\bq
\label{def_primitive_mixed}
 A_{10,3} \; = \; d \Omega_{10,3},
 \;\;\;
 A_{11,3} \; = \; d \Omega_{11,3},
 \;\;\;
 A_{12,3} \; = \; d \Omega_{12,3},
 \;\;\;
 A_{12,4} \; = \; d \Omega_{12,4}.
\eq
For the first few terms we have
\bq
\label{series_expansion}
 \Omega_{10,3}
 & = &
 \frac{1}{4} \ln\bar{q}^{\curvetwo} - \frac{1}{2}\ln\left(\bar{q}^{\curveone}-\bar{q}^{\curvetwo}\right)
 \nonumber \\
 & &
 + \left[ 5 + 27 \bar{q}^{\curvetwo} - 82 \left(\bar{q}^{\curvetwo}\right)^2  - 2310 \left(\bar{q}^{\curvetwo}\right)^3 + \dots \right] \bar{q}^{\curveone} 
 \nonumber \\
 & &
 - \frac{1}{2} \left[ 65 + 212 \bar{q}^{\curvetwo} - 11043 \left(\bar{q}^{\curvetwo}\right)^2  + \dots \right] \left(\bar{q}^{\curveone}\right)^2 
 \nonumber \\
 & &
 + \frac{1}{3} \left[ 689 - 4968 \bar{q}^{\curvetwo} + \dots \right] \left(\bar{q}^{\curveone}\right)^3 + \dots,
 \nonumber \\
 \Omega_{11,3}
 & = &
 -\sqrt{3} \left[
  \bar{q}^{\curveone} - 3 \bar{q}^{\curvetwo}
  + \frac{13}{2} \left(\bar{q}^{\curveone}\right)^2 - 12 \bar{q}^{\curveone} \bar{q}^{\curvetwo} - \frac{3}{2} \left(\bar{q}^{\curvetwo}\right)^2
  - 81 \left(\bar{q}^{\curveone}\right)^3 
 \right. \nonumber \\
 & & \left. 
  + 363 \left(\bar{q}^{\curveone}\right)^2 \bar{q}^{\curvetwo} - 438 \bar{q}^{\curveone} \left(\bar{q}^{\curvetwo}\right)^2 + 150 \left(\bar{q}^{\curvetwo}\right)^3 + \dots
 \right],
 \nonumber \\
 \Omega_{12,3}
 & = &
 \frac{1}{8} \ln\bar{q}^{\curveone} + \frac{1}{8} \ln\bar{q}^{\curvetwo} 
 - \frac{1}{4} \left[
  15 \bar{q}^{\curveone} + 15 \bar{q}^{\curvetwo}
  - \frac{111}{2} \left(\bar{q}^{\curveone}\right)^2 + 48 \bar{q}^{\curveone} \bar{q}^{\curvetwo} + \frac{141}{2} \left(\bar{q}^{\curvetwo}\right)^2
 \right. \nonumber \\
 & & \left.
  + 471 \left(\bar{q}^{\curveone}\right)^3 
  - 249 \left(\bar{q}^{\curveone}\right)^2 \bar{q}^{\curvetwo} + 174 \bar{q}^{\curveone} \left(\bar{q}^{\curvetwo}\right)^2 - 78\left(\bar{q}^{\curvetwo}\right)^3 + \dots
 \right],
 \nonumber \\
 \Omega_{12,4}
 & = &
 \frac{1}{4} \ln\bar{q}^{\curveone} - \frac{1}{2}\ln\left(\bar{q}^{\curvetwo}-\bar{q}^{\curveone}\right)
 \nonumber \\
 & &
 + \left[ 5 - 27 \bar{q}^{\curveone} + 53 \left(\bar{q}^{\curveone}\right)^2  + 552 \left(\bar{q}^{\curveone}\right)^3 + \dots \right] \bar{q}^{\curvetwo} 
 \nonumber \\
 & &
 - \frac{1}{2} \left[ -43 - 328 \bar{q}^{\curveone} + 11043 \left(\bar{q}^{\curveone}\right)^2  + \dots \right] \left(\bar{q}^{\curvetwo}\right)^2 
 \nonumber \\
 & &
 + \frac{1}{3} \left[ -526 + 20790 \bar{q}^{\curveone} + \dots \right] \left(\bar{q}^{\curvetwo}\right)^3 + \dots \; .
\eq
Note that the non-logarithmic terms of $\Omega_{10,3}$ start with $\bar{q}^{\curveone}$, e.g. there are no
$(\bar{q}^{\curveone})^0$-terms.
Similar, the non-logarithmic terms of $\Omega_{12,4}$ start with $\bar{q}^{\curvetwo}$, e.g. there are no
$(\bar{q}^{\curvetwo})^0$-terms.
In addition we have the relation
\bq
 \bar{q}^{\curvetwo} \frac{\partial}{\partial \bar{q}^{\curvetwo}} \Omega_{10,3} 
 +
 \bar{q}^{\curveone} \frac{\partial}{\partial \bar{q}^{\curveone}} \Omega_{12,4}
 & = & 0.
\eq
In section~\ref{sect:integrabilty} we will see that this relation
follows from integrability.


\subsection{Limits}
\label{sect:limits}

In this section we investigate the restrictions of the mixed entries to the hypersurfaces $x=0$, $y=0$ and $y-x=0$.
This provides additional information, which we use together with the information on integrability in section~\ref{sect:integrabilty}. 
Before we study the various limits of the connection matrix $A$, we first determine the residues
on the hypersurfaces $x=0$, $y=0$ and $y-x=0$.
The singular dlog-forms are
\bq
 d\ln\left(x\right) & = & \omega^{\mathrm{mpl}}_{s,0,4},
 \nonumber \\
 d\ln\left(y\right) & = & \omega^{\mathrm{modular}, \curveone}_{2,0},
 \nonumber \\
 d\ln\left(y-x\right) & = & \omega^{\mathrm{mpl}}_{s,t,1}-\omega^{\mathrm{mpl}}_{s,t,2}.
\eq
We write
\bq
 A & = & 
 C_x d\ln\left(x\right)
 + C_y d\ln\left(y\right)
 + C_{y-x} d\ln\left(y-x\right)
 + A^{\mathrm{reg}},
\eq
where $A^{\mathrm{reg}}$ is regular at $(x,y)=(0,0)$.
The residue matrices are given by
\bq
 C_x
 & = &
 \left( \begin{array}{rrrrrrrrrrrr}
 0 & 0 & 0 & 0 & 0 & 0 & 0 & 0 & 0 & 0 & 0 & 0 \\ 
 -1 & 1 & 0 & 0 & 0 & 0 & 0 & 0 & 0 & 0 & 0 & 0 \\ 
 0 & 0 & 0 & 0 & 0 & 0 & 0 & 0 & 0 & 0 & 0 & 0 \\ 
 0 & 0 & 0 & 0 & 0 & 0 & 0 & 0 & 0 & 0 & 0 & 0 \\ 
 0 & 0 & 0 & 0 & 0 & 0 & 0 & 0 & 0 & 0 & 0 & 0 \\ 
 0 & -1 & 0 & 0 & 0 & 0 & 0 & 0 & 0 & 0 & 0 & 0 \\ 
 0 & 0 & 0 & 0 & 0 & 0 & 0 & 0 & 0 & 0 & 0 & 0 \\ 
 0 & 1 & 0 & 0 & -\frac{1}{2} & 0 & 0 & 3 & -3 & 0 & 0 & 0 \\ 
 0 & 0 & 0 & 0 & -\frac{1}{6} & 0 & 0 & 1 & -1 & 0 & 0 & 0 \\ 
 0 & 0 & -\frac{1}{2} & 0 & 0 & 0 & 0 & 0 & 0 & 0 & 0 & 0 \\ 
 0 & -1 & 0 & 0 & \frac{1}{6} & 0 & 0 & -1 & 1 & 0 & 0 & 0 \\ 
 0 & 0 & 0 & -\frac{1}{2} & 0 & 0 & 0 & 0 & 0 & 0 & 0 & 0 \\ 
 \end{array} \right),
 \nonumber \\
 C_y
 & = &
 \left( \begin{array}{rrrrrrrrrrrr}
 0 & 0 & 0 & 0 & 0 & 0 & 0 & 0 & 0 & 0 & 0 & 0 \\ 
 0 & 0 & 0 & 0 & 0 & 0 & 0 & 0 & 0 & 0 & 0 & 0 \\ 
 0 & 0 & \frac{1}{2} & 1 & 0 & 0 & 0 & 0 & 0 & 0 & 0 & 0 \\ 
 0 & 0 & \frac{1}{4} & \frac{1}{2} & 0 & 0 & 0 & 0 & 0 & 0 & 0 & 0 \\ 
 0 & 0 & 0 & 0 & 0 & 0 & 0 & 0 & 0 & 0 & 0 & 0 \\ 
 0 & 0 & 0 & 0 & 0 & 0 & 0 & 0 & 0 & 0 & 0 & 0 \\ 
 0 & 0 & 0 & 0 & 0 & 0 & 0 & 0 & 0 & 0 & 0 & 0 \\ 
 0 & 0 & 0 & 0 & 0 & 0 & 0 & 0 & 0 & 0 & 0 & 0 \\ 
 0 & 0 & 0 & 0 & 0 & 0 & 0 & 0 & 0 & 0 & 0 & 0 \\ 
 0 & 0 & 0 & 0 & 0 & 0 & 0 & 0 & 0 & 0 & 0 & 0 \\ 
 0 & 0 & 0 & 0 & 0 & 0 & 0 & 0 & 0 & 0 & 0 & 0 \\ 
 0 & 0 & \frac{1}{8} & \frac{1}{4} & 0 & 0 & 0 & 0 & 0 & 0 & 0 & 0 \\ 
 \end{array} \right),
 \nonumber \\
 C_{y-x}
 & = &
 \left( \begin{array}{rrrrrrrrrrrr}
 0 & 0 & 0 & 0 & 0 & 0 & 0 & 0 & 0 & 0 & 0 & 0 \\ 
 0 & 0 & 0 & 0 & 0 & 0 & 0 & 0 & 0 & 0 & 0 & 0 \\ 
 0 & 0 & 0 & 0 & 0 & 0 & 0 & 0 & 0 & 0 & 0 & 0 \\ 
 0 & 0 & 0 & 0 & 0 & 0 & 0 & 0 & 0 & 0 & 0 & 0 \\ 
 0 & 0 & 0 & 0 & 0 & 0 & 0 & 0 & 0 & 0 & 0 & 0 \\ 
 0 & 0 & 0 & 0 & 0 & 0 & 0 & 0 & 0 & 0 & 0 & 0 \\ 
 0 & 0 & 0 & 0 & 0 & 0 & 0 & 0 & 0 & 0 & 0 & 0 \\ 
 0 & 0 & 0 & 0 & 0 & 0 & 0 & 0 & 0 & 0 & 0 & 0 \\ 
 0 & 0 & 0 & 0 & 0 & 0 & 0 & 0 & 0 & 0 & 0 & 0 \\ 
 0 & 0 & \frac{1}{4} & 0 & 0 & 0 & 0 & 0 & 0 & \frac{1}{2} & 0 & 1 \\ 
 0 & 0 & 0 & 0 & 0 & -1 & -2 & 0 & 1 & 0 & 1 & 0 \\ 
 0 & 0 & \frac{1}{8} & 0 & 0 & 0 & 0 & 0 & 0 & \frac{1}{4}  & 0 & \frac{1}{2}  \\ 
 \end{array} \right).
\eq
In particular we have
\begin{alignat}{4}
 A_{10,3} & = A_{10,3}^{\mathrm{reg}} && -\frac{1}{2} d\ln\left(x\right) && && + \frac{1}{4} d\ln\left(y-x\right),
 \nonumber \\
 A_{11,3} & = A_{11,3}^{\mathrm{reg}}, && && && 
 \nonumber \\
 A_{12,3} & = A_{12,3}^{\mathrm{reg}} && && + \frac{1}{8} d\ln\left(y\right) && + \frac{1}{8} d\ln\left(y-x\right),
 \nonumber \\
 A_{12,4} & = A_{12,4}^{\mathrm{reg}} && -\frac{1}{2} d\ln\left(x\right) && + \frac{1}{4} d\ln\left(y\right). &&
\end{alignat}
We would like to convert the singular dlog-forms in the variables $(x,y)$ to the variables $\bar{q}^{\curveone}$ and $\bar{q}^{\curvetwo}$. 
We have the following relations
\bq
 \lim\limits_{x\rightarrow 0}\left[ d\ln\left(x\right) - d\ln\left(1-\frac{\bar{q}^{\curvetwo}}{\bar{q}^{\curveone}}\right)\right]
 & = &
 2 d\ln\left(\psi_1\right) + d\ln\left(y\right) + d\ln\left(y-9\right) - d\ln\left(y+3\right),
 \nonumber \\
 \lim\limits_{y\rightarrow 0}\left[ d\ln\left(y\right) - d\ln\left(\bar{q}^{\curveone}\right)\right]
 & = &
 0,
 \nonumber \\
 \lim\limits_{x\rightarrow y}\left[ d\ln\left(y-x\right) - d\ln\left(\bar{q}^{\curvetwo}\right)\right]
 & = &
 d\ln\left(y-1\right) + d\ln\left(y+1\right) + 2 d\ln\left(y-3\right).
\eq
Note that the right-hand sides are not necessarily zero (see remark 1 in section~\ref{sect:special_kinematic_configuration}).
The first equation is derived as follows: In the limit $x\rightarrow 0$ we have
\bq
 \tau^{\curvetwo} & = & 
 \tau^{\curveone}
 + x \left. \left(\frac{\partial \tau^{\curvetwo}}{\partial x} \right)\right|_{x=0}
 + {\mathcal O}\left(x^2\right),
\eq
and
\bq
 \left. \frac{\partial \tau^{\curvetwo}}{\partial x}\right|_{x=0}
 & = &
 \left. \frac{W_x^{\curvetwo}}{(\psi_1^{\curvetwo})^2}\right|_{x=0}
 \; = \;
 \frac{2\pi i}{\psi_1^2} \cdot \frac{\left(y+3\right)}{y\left(9-y\right)}.
\eq
Hence
\bq
 \lim\limits_{x\rightarrow 0}\left[ \ln\left(x\right) - \ln\left(1-\frac{\bar{q}^{\curvetwo}}{\bar{q}^{\curveone}}\right)\right]
 & = &
 2 \ln\left(\frac{\psi_1}{2\pi i}\right) + \ln\left(\frac{y\left(9-y\right)}{\left(y+3\right)}\right).
\eq
The other two equations are derived in a similar way.

\subsubsection{The limit $x=0$}

We first consider the case $x=0$.
This corresponds to $x'=(1-y'^2)/(1+y'^2)$ and $\bar{q}^{\curveone}=\bar{q}^{\curvetwo}=\bar{q}$.
In this limit we have $\psi_1^{\curveone}=\psi_1^{\curvetwo}=\psi_1$, 
$\tau^{\curveone}=\tau^{\curvetwo}=\tau$,
$z^{\curvetwo}=0$ and
\bq
 F_{11,10}
 \; = \;
 - \frac{1}{3} \left(y-9\right) \frac{\psi_1}{\pi},
 & & 
 F_{12,3}
 \; = \;
 \frac{1}{144} \left(y^2-30y-27\right) \frac{\psi_1^2}{\pi^2}.
\eq
The matrix $A$ reduces to
\bq
\lefteqn{
 A - C_x d\ln\left(x\right)
 = 
 2 \pi i d\tau
}
 & & \\
 & &
 \left( \begin{array}{cccccccccccc}
 0 & 0 & 0 & 0 & 0 & 0 & 0 & 0 & 0 & 0 & 0 & 0 \\ 
 0 & 0 & 0 & 0 & 0 & 0 & 0 & 0 & 0 & 0 & 0 & 0 \\ 
 0 & 0 & -f_2 & 1 & 0 & 0 & 0 & 0 & 0 & 0 & 0 & 0 \\ 
 -\frac{1}{2} g_3 & 0 & f_4 & -f_2 & 0 & 0 & 0 & 0 & 0 & 0 & 0 & 0 \\ 
 0 & 0 & 0 & 0 & 0 & 0 & 0 & 0 & 0 & 0 & 0 & 0 \\ 
 0 & 0 & 0 & 0 & 0 & 0 & 0 & 0 & 0 & 0 & 0 & 0 \\ 
 0 & 0 & -\frac{1}{3} g_3 & 0 & 0 & 0 & 0 & 0 & 0 & 0 & 0 & 0 \\ 
 0 & 0 & 0 & 0 & 0 & 0 & 0 & 0 & 0 & 0 & 0 & 0 \\ 
 0 & 0 & 0 & 0 & 0 & 0 & 0 & 0 & 0 & 0 & 0 & 0 \\ 
 0 & 0 & \frac{1}{4} h_2 & 0 & 0 & 0 & 0 & 0 & 0 & -f_2 & 0 & 1 \\ 
 0 & 0 & -h_3 & 0 & \frac{1}{9} g_{2,1} & -k_2 & -2 k_2 & -\frac{2}{3} g_{2,1} & g_{2,0} & 0 & g_2 & 0 \\ 
 0 & -\frac{1}{4} g_3 & h_4 & \frac{1}{4} h_2 & -k_3 & -\frac{1}{6} g_3 & -\frac{1}{3} g_3 & \frac{1}{12} g_3 & -\frac{1}{4} g_3 & f_4 & 0 & -f_2 \\ 
 \end{array} \right).
 \nonumber 
\eq
All entries are modular forms of $\Gamma_1(6)$.
At modular weight $2$ we introduced the linear combinations
\bq
 f_2 & = & - \frac{1}{2} g_{2,0} + g_{2,1} + g_{2,9},
 \nonumber \\
 g_2 & = & g_{2,0} - 2 g_{2,1},
 \nonumber \\
 h_2 & = & g_{2,0} - \frac{7}{3} g_{2,1} + g_{2,9},
 \nonumber \\
 k_2 & = & g_{2,0} - \frac{4}{3} g_{2,1}.
\eq
At modular weight $3$ and $4$ we have apart from the modular forms $g_3$ and $f_4$ defined previously
in addition the modular forms
\bq
 h_3
 & = &
 \frac{1}{108} y \left(y+3\right)\left(y-9\right) \frac{\psi_1^3}{\pi^3},
 \nonumber \\
 k_3
 & = &
 \frac{1}{864} y \left(y+23\right)\left(y-9\right) \frac{\psi_1^3}{\pi^3},
 \nonumber \\
 h_4
 & = &
 \frac{1}{1728} \left(13y^4 -194y^3+432y^2-1134y+243\right) \frac{\psi_1^4}{\pi^4}.
\eq
All entries may be expressed as polynomials in
\bq
\label{def_b_basis}
 \frac{\psi_1}{\pi}
 & \mbox{and} &
 \frac{\psi_1}{\pi} y.
\eq
In order to obtain the $\bar{q}^{\curveone}$-expansion of the integration kernels 
we introduce a basis $\{e_{1,1},e_{1,2}\}$ for the modular forms of modular weight $1$ 
for the Eisenstein subspace ${\mathcal E}_1(\Gamma_1(6))$:
\bq
 e_{1,1} \; = \; E_1\left(\tau^{\curveone};\chi_{1},\chi_{-3}\right),
 & &
 e_{1,2} \; = \; E_1\left(2\tau^{\curveone};\chi_{1},\chi_{-3}\right),
\eq
where $\chi_{1}$ and $\chi_{-3}$ denote primitive Dirichlet characters with conductors $1$ and $3$, respectively.
All occurring integration kernels may be expressed as polynomials in $e_{1,1}$ and $e_{1,2}$, by expressing
$\frac{\psi_1}{\pi}$and $\frac{\psi_1}{\pi} y$
in terms of $e_{1,1}$ and $e_{1,2}$:
\bq
 \frac{\psi_1}{\pi}
 \; = \;
 2 \sqrt{3}
 \left( e_{1,1} + e_{1,2} \right),
 & &
 \frac{\psi_1}{\pi} y
 \; = \; 
 6 \sqrt{3}
 \left( e_{1,1} - e_{1,2} \right).
\eq
Alternatively, we may express these two functions in terms of the functions $g_1(z,\tau)$:
\bq
 \frac{\psi_1}{\pi}
 \; = \;
 \frac{1}{2\pi} \left( g^{(1)}(\frac{1}{3},\tau) + g^{(1)}(\frac{1}{6},\tau) \right),
 & &
 \frac{\psi_1}{\pi} y
 \; = \; 
 \frac{1}{2\pi} \left( 9 g^{(1)}(\frac{1}{3},\tau) - 3 g^{(1)}(\frac{1}{6},\tau) \right).
\eq
The four mixed entries $A_{10,3}$, $A_{11,3}$, $A_{12,3}$ and $A_{12,4}$
are given in terms of the one-forms $\omega^{\mathrm{Kronecker}}_{k}(z,\tau)$
by
\bq
 & &
 A_{10,3} + \frac{1}{2}d\ln\left(x\right)
 \; = \;
 A_{12,4} + \frac{1}{2}d\ln\left(x\right)
 \; = \;
 \omega^{\mathrm{Kronecker}}_{2}(\frac{1}{2},\tau)
 + 2 \omega^{\mathrm{Kronecker}}_{2}(\frac{1}{3},\tau),
 \nonumber \\
 & &
 A_{11,3}
 \; = \; 
 -2 i \omega^{\mathrm{Kronecker}}_{3}(\frac{1}{3},\tau)
 + 4 i \omega^{\mathrm{Kronecker}}_{3}(\frac{1}{6},\tau),
 \\
 & & 
 A_{12,3}
 \; = \;  
 -11  \omega^{\mathrm{Kronecker}}_{4}(0,\tau)
 -28  \omega^{\mathrm{Kronecker}}_{4}(\frac{1}{2},\tau)
 -    \omega^{\mathrm{Kronecker}}_{4}(\frac{1}{3},\tau)
 - 20 \omega^{\mathrm{Kronecker}}_{4}(\frac{1}{6},\tau).
 \nonumber
\eq

\subsubsection{The limit $x=y$}

As a second limiting case we consider the case $x=y$.
The case $x=y$ corresponds to $x'=1$ and $\bar{q}^{\curvetwo}=0$.
In this case we have
\bq
 \frac{\psi_1^{\curvetwo}}{\pi} \; = \; \frac{2}{\sqrt{\left(1+y\right)\left(3-y\right)}},
 & &
 \lim\limits_{x \rightarrow y} \frac{\partial_y \psi_1^{\curvetwo}}{\pi} \; = \; \frac{2\left(3+3y-3y^2+y^3\right)}{\sqrt{\left(1+y\right)\left(3-y\right)} \left(1-y^2\right)\left(3-y\right)^2}.
\eq
On the left-hand side
of the second equation we first take the derivative with respect to $y$ and then the limit $x \rightarrow y$.
The functions $F_{11,10}$ and $F_{12,3}$ reduce to
\bq
 F_{11,10} \; = \; 2 \sqrt{\frac{3-y}{1+y}},
 & &
 F_{12,3} \; = \; -\frac{1}{4}.
\eq
In this limit we have
\bq
 z^{\curvetwo}
 & = &
 \frac{1}{2\pi i} \ln\left(\frac{r_3}{2} \left( y-1 - i \sqrt{\left(1+y\right)\left(3-y\right)}\right)\right),
 \nonumber \\
 y & = & \frac{r_3}{\bar{w}^{\curvetwo}} \left[ 1 + \frac{\bar{w}^{\curvetwo}}{r_3} + \left(\frac{\bar{w}^{\curvetwo}}{r_3}\right)^2 \right],
\eq
where $r_3=\exp(2\pi i/3)$ denotes a third root of unity.
We have
\bq
 A_{10,3} - \frac{1}{4} d\ln\left(y-x\right)
 & = &
 - \left[ \frac{\left(9-6y+5y^2\right)}{12 y \sqrt{\left(1+y\right)\left(3-y\right)}} \frac{\psi_1^{\curveone}}{\pi}
 + \frac{1}{4\left(y-1\right)}
 + \frac{1}{4\left(y+1\right)}
 + \frac{1}{2\left(y-3\right)}
 \right] dy,
 \nonumber \\
 A_{11,3}
 & = &
 - \frac{2}{3} \omega^{\mathrm{modular}, \curveone}_{3} + \frac{dy}{\sqrt{\left(1+y\right)\left(3-y\right)}},
 \nonumber \\
 A_{12,3} - \frac{1}{8} d\ln\left(y-x\right)
 & = &
 \left[ \frac{\left(9-6y+5y^2\right)}{24 y \sqrt{\left(1+y\right)\left(3-y\right)}} \frac{\psi_1^{\curveone}}{\pi} 
 - \frac{1}{8y}
 + \frac{1}{8\left(y-1\right)}
 - \frac{1}{8\left(y+1\right)}
 - \frac{1}{2\left(y-3\right)}
 \right. \nonumber \\
 & & \left.
 + \frac{1}{4\left(y-9\right)}
 \right] dy,
 \nonumber \\
 A_{12,4}
 & = &
 - \frac{1}{4} \omega_0^{\mathrm{modular}, \curveone}.
\eq

\subsubsection{The limit $y=0$}

As a third limiting case we consider the case $y=0$.
The case $y=0$ corresponds to $y'=0$ and $\bar{q}^{\curveone}=0$.
In this case we
\bq
 \frac{\psi_1^{\curveone}}{\pi} \; = \; \frac{2}{3} \sqrt{3},
 & &
 \frac{\partial_y \psi_1^{\curveone}}{\pi} \; = \; \frac{2}{9} \sqrt{3}.
\eq
$F_{12,3}$ reduces to
\bq
 F_{12,3}
 & = &
 \frac{1}{4} - \frac{\sqrt{3}}{12} \frac{\left(9-14x+9x^2\right)}{\left(3-4x+3x^2\right)} \frac{\psi_1^{\curvetwo}}{\pi}.
\eq
We have
\bq
 A_{10,3}
 & = &
 - \frac{1}{4} \omega_0^{\mathrm{modular}, \curvetwo},
 \nonumber \\
 A_{11,3}
 & = &
 i \omega^{\mathrm{Kronecker}, \curvetwo}_{1},
 \nonumber \\
 A_{12,3} - \frac{1}{8} d\ln\left(y\right)
 & = &
 \frac{\sqrt{3}}{8} \left( \frac{1}{x} - \frac{2}{x-1} \right) \frac{\psi_1^{\curvetwo}}{\pi} dx
 - \frac{1}{4} A_{10,10},
 \nonumber \\
 A_{12,4} - \frac{1}{4} d\ln\left(y\right)
 & = &
 - \frac{\sqrt{3}}{4} \left( \frac{1}{x} - \frac{2}{x-1} \right) \frac{\psi_1^{\curvetwo}}{\pi} dx.
\eq


\subsection{Integrability}
\label{sect:integrabilty}

In this section we explore the constraints from integrability.
 
The connection matrix $A$ appearing in the differential equation~(\ref{differential_equation_eps_form}) is flat (or integrable):
\bq
 d A - A \wedge A & = & 0.
\eq
As $A$ is proportional to $\eps$, it follows that $dA$ is proportional to $\eps$ while
$A \wedge A$ is proportional to $\eps^2$.
This implies the two separate equations
\bq
 dA \; = \; 0,
 & &
 A \wedge A \; = \; 0.
\eq
The equation $dA=0$ states that all entries of $A$ are closed differential one-forms.
If an entry can be written as  double series in $\bar{q}^{\curveone}$ and $\bar{q}^{\curvetwo}$
\bq
 A_{i,j} & = &
 \sum\limits_{k,l} c_{k,l}^{\curveone} \left( \bar{q}^{\curveone} \right)^k \left( \bar{q}^{\curvetwo} \right)^l  \frac{d\bar{q}^{\curveone}}{\bar{q}^{\curveone}}
 +
 \sum\limits_{k,l} c_{k,l}^{\curvetwo} \left( \bar{q}^{\curveone} \right)^k \left( \bar{q}^{\curvetwo} \right)^l  \frac{d\bar{q}^{\curvetwo}}{\bar{q}^{\curvetwo}},
\eq
the closedness condition implies for the coefficients $c_{k,l}^{\curveone}$ and $c_{k,l}^{\curvetwo}$
\bq 
 l c_{k,l}^{\curveone} - k c_{k,l}^{\curvetwo} & = & 0.
\eq
This implies in particular 
\bq
 c_{0,l}^{\curveone} & = &  0
 \;\;\;\mbox{for}\;\;\;
 l \; \neq \; 0,
 \nonumber \\
 c_{k,0}^{\curvetwo} & = & 0
 \;\;\;\mbox{for}\;\;\;
 k \; \neq \; 0.
\eq
The equation $A \wedge A = 0$ implies seven relations for the mixed entries $A_{10,3}$, $A_{11,3}$, $A_{12,3}$ and $A_{12,4}$.
The relations are
\bq
\label{integrability_relations}
\lefteqn{
 A_{10,3} \wedge \omega_0^{\mathrm{modular}, \curveone} - A_{12,4} \wedge \omega_0^{\mathrm{modular}, \curvetwo} = 0,
} & &
 \nonumber \\
\lefteqn{
 A_{11,3} \wedge \omega_0^{\mathrm{modular}, \curveone} + 4 i A_{12,4} \wedge \omega_1^{\mathrm{Kronecker}, \curvetwo} = 0,
} & &
 \nonumber \\
\lefteqn{
 A_{12,3} \wedge \omega_0^{\mathrm{modular}, \curveone} - A_{12,4} \wedge \left( A_{12,12} - A_{4,4} \right) = 0,
} & &
 \nonumber \\
\lefteqn{
 A_{10,3} \wedge \left( A_{3,3} - A_{10,10} \right) 
 + \frac{3}{2} i A_{11,3} \wedge \omega_1^{\mathrm{Kronecker}, \curvetwo}
 - A_{12,3} \wedge \omega_0^{\mathrm{modular}, \curvetwo}
 } & & \nonumber \\
 & &
 + \frac{2}{3} i \omega_3^{\mathrm{modular}, \curveone} \wedge \omega_1^{\mathrm{Kronecker}, \curvetwo}
 = 0,
 \hspace*{50mm}
 \nonumber \\
\lefteqn{
 A_{10,3} \wedge A_{11,10} + A_{11,3} \wedge \left( A_{11,11} - A_{3,3} \right) - 4 i A_{12,3} \wedge \omega_1^{\mathrm{Kronecker}, \curvetwo}
} & & \nonumber \\
 & & 
 - \frac{1}{3} \omega_3^{\mathrm{modular}, \curveone} \wedge A_{11,7} 
 = 0,
 \nonumber \\
\lefteqn{
 A_{10,3} \wedge A_{12,10} + A_{11,3} \wedge A_{12,11} + A_{12,3} \wedge \left( A_{12,12} - A_{3,3} \right) 
 - A_{12,4} \wedge \omega_4^{\mathrm{modular}, \curveone}
} & & \nonumber \\
 &&
 - \frac{1}{3} \omega_3^{\mathrm{modular}, \curveone} \wedge A_{12,7} 
 = 0,
 \nonumber \\
\lefteqn{
 A_{12,4} \wedge \omega_3^{\mathrm{modular}, \curveone}
 - 2 A_{12,2} \wedge A_{2,1} = 0.
} & &
\eq
If we write
\bq
 A & = & 
 A^{\curveone} 2 \pi i d\tau^{\curveone}
 +
 A^{\curvetwo} 2 \pi i d\tau^{\curvetwo},
\eq
integrability alone allows us to express seven of the eight functions
\bq
 A_{10,3}^{\curveone}, \; A_{10,3}^{\curvetwo}, \;
 A_{11,3}^{\curveone}, \; A_{11,3}^{\curvetwo}, \;
 A_{12,3}^{\curveone}, \; A_{12,3}^{\curvetwo}, \;
 A_{12,4}^{\curveone}, \; A_{12,4}^{\curvetwo}
\eq
in terms of one function to be determined by other means.

With the additional information on the limits and the closedness we may determine
all functions in terms of the variables $\tau^{\curveone}, \tau^{\curvetwo}, z^{\curvetwo}$ as follows:
We first note that the first six equations of eq.~(\ref{integrability_relations})
define a linear system of equations for the six functions
\bq
 A_{10,3}^{\curveone}, \; A_{10,3}^{\curvetwo}, \;
 A_{11,3}^{\curveone}, \; A_{11,3}^{\curvetwo}, \;
 A_{12,3}^{\curveone}, \; A_{12,3}^{\curvetwo}.
\eq
Solving this system allows us to express these six functions 
in terms of 
\bq
 A_{12,4}^{\curveone}, \; A_{12,4}^{\curvetwo}
\eq
and other (already known) functions.
It is therefore sufficient to focus on $A_{12,4}^{\curveone}$ and $A_{12,4}^{\curvetwo}$.
The latter we get directly from the seventh equation of eq.~(\ref{integrability_relations}), the former 
then from the closedness property and the limit $x=y$.
Let's see how this works out in detail:
We write
\bq
 A_{12,2} \wedge A_{2,1}
 & = &
 - \frac{i}{2} 
 \left[ \omega^{\mathrm{Kronecker},\frac{1}{2}}_{3}\left(3z^{\curvetwo},\tau^{\curvetwo}\right)
        - 3 \omega^{\mathrm{Kronecker},\frac{1}{2}}_{3}\left(z^{\curvetwo}+\frac{2}{3},\tau^{\curvetwo}\right)
 \right]
 \nonumber \\
 & &
 \wedge
 \left[ \omega^{\mathrm{Kronecker},\frac{1}{2}}_{2}\left(3z^{\curvetwo},\tau^{\curvetwo}\right)
        - \omega^{\mathrm{Kronecker},\frac{1}{2}}_{2}\left(z^{\curvetwo}+\frac{2}{3},\tau^{\curvetwo}\right)
\right] 
 \nonumber \\
 & = &
 - \frac{i}{2} 
 H_4\left(z^{\curvetwo},\tau^{\curvetwo}\right) \left(2\pi i dz^{\curvetwo} \right) \wedge \left(2\pi i d\tau^{\curvetwo} \right),
\eq
where $H_4$ is given by
\bq
\lefteqn{
 H_4\left(z^{\curvetwo},\tau^{\curvetwo}\right)
 = 
 \frac{1}{\left(2\pi i\right)^4}
 } & & \\
 & &
 \left\{
  \left[ h^{(2)}\left(3z^{\curvetwo},\tau^{\curvetwo}\right) - 3 h^{(2)}\left(z^{\curvetwo}+\frac{2}{3},\tau^{\curvetwo}\right) \right]
  \left[ h^{(2)}\left(3z^{\curvetwo},\tau^{\curvetwo}\right) - h^{(2)}\left(z^{\curvetwo}+\frac{2}{3},\tau^{\curvetwo}\right) \right]
 \right. \nonumber \\
 & & \left.
  -
  \left[ h^{(3)}\left(3z^{\curvetwo},\tau^{\curvetwo}\right) - 3 h^{(3)}\left(z^{\curvetwo}+\frac{2}{3},\tau^{\curvetwo}\right) \right]
  \left[ h^{(1)}\left(3z^{\curvetwo},\tau^{\curvetwo}\right) - h^{(1)}\left(z^{\curvetwo}+\frac{2}{3},\tau^{\curvetwo}\right) \right]
 \right\}.
 \nonumber
\eq
$H_4$ is of modular weight four with respect to curve $\curvetwo$.
Then
\bq
 A_{12,4}^{\curvetwo}
 & = &
 i \frac{H_4\left(z^{\curvetwo},\tau^{\curvetwo}\right)}{g_3\left(\tau^{\curveone}\right)} 
 \frac{\partial z^{\curvetwo}(\tau^{\curveone},\tau^{\curvetwo})}{\partial \tau^{\curveone}}.
\eq
Integration in $\tau^{\curvetwo}$ gives $\Omega_{12,4}$ up to a function depending on $\tau^{\curveone}$, 
but independent of $\tau^{\curvetwo}$.
However, this function we know from the limit $x=y$ (and the expansion in eq.~(\ref{series_expansion})).
We find
\bq
 \Omega_{12,4} 
 & = &
 \frac{1}{4} \ln \bar{q}^{\curveone}
 +
 \frac{i}{g_3\left(\tau^{\curveone}\right)}
 \int \frac{d\bar{q}^{\curvetwo}}{\bar{q}^{\curvetwo}}
 H_4\left(z^{\curvetwo},\tau^{\curvetwo}\right)
 \frac{\partial z^{\curvetwo}(\tau^{\curveone},\tau^{\curvetwo})}{\partial \tau^{\curveone}}.
\eq
In the integrand $z^{\curvetwo}$ is viewed as a function of $\tau^{\curveone}$ and $\tau^{\curvetwo}$:
\bq
 z^{\curvetwo} & = & z^{\curvetwo}\left(\tau^{\curveone},\tau^{\curvetwo}\right).
\eq
$A_{12,4}^{\curveone}$ is then given by
\bq
 A_{12,4}^{\curveone}
 & = &
 \left( \bar{q}^{\curveone} \frac{\partial \Omega_{12,4}}{\partial \bar{q}^{\curveone}} \right) \cdot 2 \pi i d\tau^{\curveone}.
\eq


\section{Conclusions}
\label{sect:conclusions}

We studied a two-loop Feynman integral with four external legs and one internal mass,
depending on two kinematic variables.
This Feynman integral has two elliptic curves associated to it: One elliptic curve (curve $\curvetwo$ in our notation)
is associated to the maximal cut of the top sector, the second elliptic curve (curve $\curveone$ in our notation)
is associated to the sunrise sub-topology.
For generic kinematic variables the two curves are not isogenic (and hence not isomorphic).
We studied the differential equation for this family of Feynman integrals.
Our main results are threefold:
We first showed that the differential equation can be transformed to an $\eps$-form.
To the best of our knowledge, this is the first time this has been achieved for a Feynman integral
beyond the ones evaluating to multiple polylogarithms or depending on a single elliptic curve.
This result supports the conjecture that an $\eps$-form can be reached for any Feynman integral.

We then studied the entries of the differential equation, and here in particular the ones
giving the derivatives of the three master integrals in the top sector.
We found that most of these entries depend only on curve $\curvetwo$, but not on curve $\curveone$.
These entries can be expressed in terms of differential one-forms already encountered in 
the unequal-mass sunrise integral.
This shows the universality of these differential one-forms.
This is our second main result.

There are four entries, which depend on both elliptic curves. We studied them in detail.
In particular we expressed them in the natural coordinates (from a mathematical point of view)
$\tau^{\curveone}$ and $(\tau^{\curvetwo},z^{\curvetwo})$.
The former is a coordinate on the moduli space ${\mathcal M}_{1,1}$ associated with curve $\curveone$,
the latter two are coordinates on the moduli space ${\mathcal M}_{1,2}$ associated with curve $\curvetwo$.
As the Feynman integral under consideration depends only on two kinematic variables, we may express
one variable from the set $(\tau^{\curveone},\tau^{\curvetwo},z^{\curvetwo})$ in terms of the other two.
A representation in terms of three variables $(\tau^{\curveone},\tau^{\curvetwo},z^{\curvetwo})$ 
(together with the relation among the variables)
is therefore not unique.
However, integrability gives us a natural representation of the mixed entries in terms of these three variables,
which makes the modular transformation properties with respect to the two elliptic curves transparent.
This is our third main result.

We expect that the patterns found in this Feynman integral carry over to more complicated Feynman integrals.


\begin{appendix}

\section{The mixed entries in $(x,y)$-coordinates}
\label{sect:mixed_xy}

The four mixed entries in the $(x,y)$-coordinates read
\bq
 A_{10,3}
 & = &
 \left[ 
  - \frac{\left(1-x\right) N_1}{6x\left(1+x\right)P_1} \frac{\psi_1^{\curveone}}{\psi_1^{\curvetwo}}
  + \frac{4\left(1-x\right)N_2}{\left(1+x\right)\left(y-x\right)\left(1-xy\right)P_2} \left(\frac{\pi}{\psi_1^{\curvetwo}}\right)^2 F_{12,3}
 \right] dx
 \nonumber \\
 & &
  - \frac{4\left(1-x\right)^2 P_1}{\left(1-y\right)\left(y-x\right)\left(1-xy\right)P_2} \left(\frac{\pi}{\psi_1^{\curvetwo}}\right)^2 F_{12,3} dy,
 \nonumber \\
 A_{11,3}
 & = &
 \left[
  \frac{\left(1-x\right) N_3}{6x\left(1+x\right) P_1} \frac{\psi_1^{\curveone}}{\pi}
  - \frac{\left(1-x\right)N_1}{6x\left(1+x\right) P_1} \frac{\psi_1^{\curveone}}{\psi_1^{\curvetwo}} F_{11,10}
  - \frac{4 \left(1-x\right)\left(1+y\right)}{\left(1+x\right)\left(y-x\right)\left(1-xy\right)} \frac{\pi}{\psi_1^{\curvetwo}} F_{12,3}
 \right. \nonumber \\
 & & \left.
  + \frac{4 \left(1-x\right)N_2}{\left(1+x\right)\left(y-x\right)\left(1-xy\right) P_2} \left(\frac{\pi}{\psi_1^{\curvetwo}}\right)^2 F_{11,10} F_{12,3}
 \right] dx
 \nonumber \\
 & & 
 +
 \left[
  \frac{\left(1-x\right)^2 \left(3+y\right)}{3\left(1-y\right)P_1} \frac{\psi_1^{\curveone}}{\pi}
  + \frac{4\left(1-x\right)^2}{\left(1-y\right)\left(y-x\right)\left(1-xy\right)} \frac{\pi}{\psi_1^{\curvetwo}} F_{12,3}
 \right. \nonumber \\
 & & \left.
  - \frac{4\left(1-x\right)^2P_1}{\left(1-y\right)\left(y-x\right)\left(1-xy\right) P_2} \left(\frac{\pi}{\psi_1^{\curvetwo}}\right)^2 F_{11,10} F_{12,3}
 \right] dy,
 \nonumber \\
 A_{12,3}
 & = &
 \left[
  \frac{N_4}{96 x \left(1-x\right)\left(1+x\right) P_1} \frac{\psi_1^{\curveone}}{\pi} \frac{\psi_1^{\curvetwo}}{\pi}
  - \frac{\left(1-x\right)N_3}{16 x \left(1+x\right) P_1} \frac{\psi_1^{\curveone}}{\pi} F_{11,10}
 \right. \nonumber \\
 & & \left.
  - \frac{N_5}{4 \left(1-x\right) \left(1+x\right) \left(y-x\right)\left(1-xy\right) P_2} F_{12,3}
  + \frac{3\left(1-x\right)\left(1+y\right)}{2\left(1+x\right)\left(y-x\right)\left(1-xy\right)} \frac{\pi}{\psi_1^{\curvetwo}} F_{11,10} F_{12,3}
 \right. \nonumber \\
 & & \left.
  + \frac{\left(1-x\right)N_1}{32 x \left(1+x\right) P_1} \frac{\psi_1^{\curveone}}{\psi_1^{\curvetwo}} F_{11,10}^2
  - \frac{3\left(1-x\right) N_2}{4\left(1+x\right)\left(y-x\right)\left(1-xy\right) P_2} \left(\frac{\pi}{\psi_1^{\curvetwo}}\right)^2 F_{11,10}^2 F_{12,3}
 \right] dx
 \nonumber \\
 & &
 + \left[
  \frac{\left(3+y\right) N_6}{48 y \left(1-y\right)\left(9-y\right) P_1} \frac{\psi_1^{\curveone}}{\pi} \frac{\psi_1^{\curvetwo}}{\pi}
  - \frac{N_7}{24 \left(1-y\right) P_1} \frac{\psi_1^{\curveone}}{\pi} F_{11,10}
 \right. \nonumber \\
 & & \left.
  - \frac{N_8}{4 y \left(1-y\right)\left(9-y\right)\left(y-x\right)\left(1-xy\right) P_2} F_{12,3}
  - \frac{3\left(1-x\right)^2}{2\left(1-y\right)\left(y-x\right)\left(1-xy\right)} \frac{\pi}{\psi_1^{\curvetwo}} F_{11,10} F_{12,3}
 \right. \nonumber \\
 & & \left.
  + \frac{3\left(1-x\right)^2 P_1}{4\left(1-y\right)\left(y-x\right)\left(1-xy\right) P_2} \left(\frac{\pi}{\psi_1^{\curvetwo}}\right)^2 F_{11,10}^2 F_{12,3}
 \right] dy,
 \nonumber \\
 A_{12,4}
 & = &
 - \frac{\left(1+x\right)}{2x\left(1-x\right)} \frac{\psi_1^{\curvetwo}}{\psi_1^{\curveone}} dx
 \nonumber \\
 & &
 + \left[ 
  \frac{N_{9}}{2y\left(1-y\right)\left(9-y\right)P_1} \frac{\psi_1^{\curvetwo}}{\psi_1^{\curveone}}
  + \frac{12}{y\left(1-y\right)\left(9-y\right)} \left(\frac{\pi}{\psi_1^{\curveone}}\right)^2 F_{12,3}
 \right] dy.
\eq
Polynomials appearing in the denominator (and possibly in the numerator) are
\bq
 P_1
 & = &
 3-4\,x+3\,{x}^{2}-2\,yx,
 \nonumber \\
 P_2
 & = &
 9-14\,x-y+9\,{x}^{2}-2\,yx-y{x}^{2}.
\eq
Polynomials appearing in the numerator are
\bq
 N_{1}
 & = & 
 9-6\,x+9\,{x}^{2}-4\,yx,
 \nonumber \\
 N_{2}
 & = &
 3-2\,x+y+3\,{x}^{2}-6\,yx+y{x}^{2},
 \nonumber \\
 N_{3}
 & = & 
 27-42\,x-3\,y+27\,{x}^{2}-22\,yx-3\,y{x}^{2},
 \nonumber \\
 N_{4}
 & = &
 243-912\,x-54\,y+1530\,{x}^{2}+8\,yx+3\,{y}^{2}-912\,{x}^{3}-4\,y{x}^{2}+243\,{x}^{4}+8\,y{x}^{3}-134\,{y}^{2}{x}^{2}
 \nonumber \\
 & & 
 +8\,{y}^{3}x-54\,y{x}^{4}+16\,{y}^{3}{x}^{2}+3\,{y}^{2}{x}^{4}+8\,{y}^{3}{x}^{3},
 \nonumber \\
 N_{5}
 & = &
 63-316\,x+103\,y+522\,{x}^{2}-268\,yx-43\,{y}^{2}-316\,{x}^{3}+282\,y{x}^{2}+76\,{y}^{2}x+5\,{y}^{3}+63\,{x}^{4}
 \nonumber \\
 & &
 -268\,y{x}^{3}-18\,{y}^{2}{x}^{2}-4\,{y}^{3}x+103\,y{x}^{4}+76\,{y}^{2}{x}^{3}-18\,{y}^{3}{x}^{2}-43\,{y}^{2}{x}^{4}-4\,{y}^{3}{x}^{3}+5\,{y}^{3}{x}^{4},
 \nonumber \\
 N_{6}
 & = & 
 81-126\,x-27\,y+81\,{x}^{2}-134\,yx-93\,{y}^{2}-27\,y{x}^{2}+326\,{y}^{2}x+7\,{y}^{3}-93\,{y}^{2}{x}^{2}-2\,{y}^{3}x
 \nonumber \\
 & &
 +7\,{y}^{3}{x}^{2},
 \nonumber \\
 N_{7}
 & = &
 21-34\,x-9\,y+21\,{x}^{2}+2\,yx-9\,y{x}^{2}+8\,{y}^{2}x,
 \nonumber \\
 N_{8}
 & = & 
 162\,x-729\,y-252\,{x}^{2}+2556\,yx+243\,{y}^{2}+162\,{x}^{3}-3862\,y{x}^{2}-932\,{y}^{2}x-27\,{y}^{3}
 \nonumber \\
 & &
 +2556\,y{x}^{3}+1554\,{y}^{2}{x}^{2}+336\,{y}^{3}x+{y}^{4}-729\,y{x}^{4}-932\,{y}^{2}{x}^{3}-618\,{y}^{3}{x}^{2}-78\,{y}^{4}x
 \nonumber \\
 & & 
 +243\,{y}^{2}{x}^{4}+336\,{y}^{3}{x}^{3}+98\,{y}^{4}{x}^{2}+4\,{y}^{5}x-27\,{y}^{3}{x}^{4}-78\,{y}^{4}{x}^{3}+8\,{y}^{5}{x}^{2}+{y}^{4}{x}^{4}+4\,{y}^{5}{x}^{3},
 \nonumber \\
 N_{9}
 & = &
 27-42\,x+30\,y+27\,{x}^{2}-52\,yx-{y}^{2}+30\,y{x}^{2}-18\,{y}^{2}x-{y}^{2}{x}^{2}.
\eq


\section{Differential one-forms}
\label{sect:dlog_forms}

In this appendix we give the expressions of the dlog-forms in terms of the coefficients of the Kronecker function.
\bq
 \omega^{\mathrm{mpl}}_{s,0}
 & = &
  - \omega^{\mathrm{Kronecker}}_{2}\left(3z^{\curvetwo},\tau^{\curvetwo}\right)
  + 2 \omega^{\mathrm{Kronecker}}_{2}\left(2z^{\curvetwo}+\frac{5}{6},\tau^{\curvetwo}\right)
  + \omega^{\mathrm{Kronecker}}_{2}\left(z^{\curvetwo}+\frac{2}{3},\tau^{\curvetwo}\right)
 \nonumber \\
 & &
  - 2 \omega^{\mathrm{modular}, \curvetwo}_{2},
 \nonumber \\
 \omega^{\mathrm{mpl}}_{s,4}
 & = &
  - \omega^{\mathrm{Kronecker}}_{2}\left(3z^{\curvetwo},\tau^{\curvetwo}\right)
  + 2 \omega^{\mathrm{Kronecker}}_{2}\left(2z^{\curvetwo}+\frac{1}{3},\tau^{\curvetwo}\right)
  - \omega^{\mathrm{Kronecker}}_{2}\left(z^{\curvetwo}+\frac{2}{3},\tau^{\curvetwo}\right)
 \nonumber \\
 & &
  + 2 \omega^{\mathrm{Kronecker}}_{2}\left(z^{\curvetwo}+\frac{1}{6},\tau^{\curvetwo}\right)
  - 2 \omega^{\mathrm{modular}, \curvetwo}_{2},
 \nonumber \\
 \omega^{\mathrm{mpl}}_{s,0,4}
 & = &
  \omega^{\mathrm{Kronecker},\frac{1}{2}}_{2}\left(3z^{\curvetwo},\tau^{\curvetwo}\right)
  - \omega^{\mathrm{Kronecker},\frac{1}{2}}_{2}\left(z^{\curvetwo}+\frac{2}{3},\tau^{\curvetwo}\right),
 \nonumber \\
 \omega^{\mathrm{mpl}}_{t,1}
 & = &
  - 2 \omega^{\mathrm{Kronecker}}_{2}\left(2z^{\curvetwo}+\frac{1}{3},\tau^{\curvetwo}\right)
  + \omega^{\mathrm{Kronecker}}_{2}\left(2z^{\curvetwo}+\frac{5}{6},\tau^{\curvetwo}\right)
 \nonumber \\
 & &
  + 2 \omega^{\mathrm{Kronecker}}_{2}\left(z^{\curvetwo}+\frac{2}{3},\tau^{\curvetwo}\right)
  + 2 \omega^{\mathrm{Kronecker}}_{2}\left(z^{\curvetwo}+\frac{1}{6},\tau^{\curvetwo}\right)
  - 3 \omega^{\mathrm{modular}, \curvetwo}_{2},
 \nonumber \\
 \omega^{\mathrm{mpl}}_{s,t,1}
 & = &
  - \omega^{\mathrm{Kronecker}}_{2}\left(3z^{\curvetwo},\tau^{\curvetwo}\right)
  + \omega^{\mathrm{Kronecker},\frac{1}{2}}_{2}\left(3z^{\curvetwo},\tau^{\curvetwo}\right)
  - 4\omega^{\mathrm{Kronecker}}_{2}\left(2z^{\curvetwo}+\frac{1}{3},\tau^{\curvetwo}\right)
 \nonumber \\
 & &
  + 2 \omega^{\mathrm{Kronecker}}_{2}\left(2z^{\curvetwo}+\frac{5}{6},\tau^{\curvetwo}\right)
  + 9 \omega^{\mathrm{Kronecker}}_{2}\left(z^{\curvetwo}+\frac{2}{3},\tau^{\curvetwo}\right)
 \nonumber \\
 & &
  + 8 \omega^{\mathrm{Kronecker}}_{2}\left(z^{\curvetwo}+\frac{1}{6},\tau^{\curvetwo}\right)
  - \omega^{\mathrm{Kronecker},\frac{1}{2}}_{2}\left(z^{\curvetwo}+\frac{2}{3},\tau^{\curvetwo}\right)
  - 2 \omega^{\mathrm{modular}, \curvetwo}_{2},
 \nonumber \\
 \omega^{\mathrm{mpl}}_{s,t,2}
 & = &
  - \frac{1}{2} \omega^{\mathrm{Kronecker}}_{2}\left(3z^{\curvetwo},\tau^{\curvetwo}\right)
  + \frac{1}{2} \omega^{\mathrm{Kronecker},\frac{1}{2}}_{2}\left(3z^{\curvetwo},\tau^{\curvetwo}\right)
 \nonumber \\
 & &
  - 2 \omega^{\mathrm{Kronecker}}_{2}\left(2z^{\curvetwo}+\frac{1}{3},\tau^{\curvetwo}\right)
  + \omega^{\mathrm{Kronecker}}_{2}\left(2z^{\curvetwo}+\frac{5}{6},\tau^{\curvetwo}\right)
 \nonumber \\
 & &
  + \frac{9}{2} \omega^{\mathrm{Kronecker}}_{2}\left(z^{\curvetwo}+\frac{2}{3},\tau^{\curvetwo}\right)
  + 4 \omega^{\mathrm{Kronecker}}_{2}\left(z^{\curvetwo}+\frac{1}{6},\tau^{\curvetwo}\right)
 \nonumber \\
 & &
  - \frac{5}{2} \omega^{\mathrm{Kronecker},\frac{1}{2}}_{2}\left(z^{\curvetwo}+\frac{2}{3},\tau^{\curvetwo}\right)
  - \omega^{\mathrm{modular}, \curvetwo}_{2}.
\eq

\end{appendix}

{\footnotesize
\bibliography{/home/stefanw/notes/biblio}
\bibliographystyle{/home/stefanw/latex-style/h-physrev5}
}

\end{document}